\documentclass[a4paper,11pt]{article}

\usepackage{jheppub} 
\usepackage{lineno}
\arxivnumber{2502.07664} 

\usepackage{tikz}
\usepackage{color}
\usepackage{amsmath}
\usepackage{amssymb}
\usepackage{amsthm}
\usepackage{dsfont}
\usepackage{mathrsfs}
\usepackage{physics}
\usepackage{textcomp, gensymb}
\usepackage{slashed}
\usepackage{braket}
\usetikzlibrary{calc}
\usepackage{array}
\usepackage{subfiles}
\usepackage{diagbox}
\usepackage{multirow}
\usepackage[all,cmtip]{xy}
\usepackage{enumitem}
\usepackage{scalerel}
\setenumerate[1]{itemsep=2pt,partopsep=0pt,parsep=\parskip,topsep=1pt}
\setitemize[1]{itemsep=2pt,partopsep=0pt,parsep=\parskip,topsep=1pt}

\let\ket\relax 

\usepackage{mathtools} 
\DeclarePairedDelimiter{\ket}{\lvert}{\rangle}

\makeatletter
\newcommand{\vast}{\bBigg@{4}}
\newcommand{\Vast}{\bBigg@{5}}
\makeatother

\makeatletter
\newcommand{\oset}[3][0ex]{%
  \mathrel{\mathop{#3}\limits^{
    \vbox to#1{\kern-2\ex@
    \hbox{$\scriptstyle#2$}\vss}}}}
\makeatother

\newcommand{\AAeta}{A_1\hspace{-0.5ex}\oset[-0.25ex]{\eta}{-}\hspace{-0.5ex}A_2}

\newcommand{\Z}{\mathbb{Z}}

\title{\boldmath Classification of Gapped Domain Walls of Topological Orders in 2+1 dimensions: A Levin-Wen Model Realization }

\author[a]{Yanyan Chen}
\author[a]{Siyuan Wang}
\author[a]{Yu Zhao}
\author[c]{Yuting Hu}
\author[a,b,1]{Yidun Wan\note{Corresponding author}}
\affiliation[a]{State Key Laboratory of Surface Physics, Center for Astronomy and Astrophysics, Department of Physics, Center for Field Theory and Particle Physics, and Institute for Nanoelectronic devices and Quantum Computing, Fudan University, 2005 Songhu Road, Shanghai 200433, China}
\affiliation[b]{Shanghai Research Center for Quantum Sciences, 99 Xiupu Road, Shanghai 201315, China}
\affiliation[b]{Hefei National Laboratory, Hefei 230088, China}
\affiliation[c]{School of Physics, Hangzhou Normal University, Hangzhou 311121, China}
\emailAdd{yanyanchen235@gmail.com, siyuanwang18@fudan.edu.cn,\\  yuzhao20@fudan.edu.cn, yuting.phys@gmail.com, ydwan@fudan.edu.cn}
 
\date{April 2023}

\abstract{This paper introduces a novel systematic construction of gapped domain walls (GDWs) within the Levin-Wen (LW) model.
By gluing two LW models along their open sides in a compatible way, we achieve a complete GDW classification by subsets of bulk input data, which encompass the classifications in terms of bimodule categories. A generalized bimodule structure is introduced to capture domain-wall excitations. Furthermore, we demonstrate that folding along any GDW yields a gapped boundary (GB) described by a Frobenius algebra of the input UFC for the folded model, thus bridging our GDW classification and the GB classification in \cite{hu2018boundary}.}

\begin{document}

\maketitle
\flushbottom

\section{Introduction}

Topological phases of matter go beyond the conventional Landau-Ginzburg symmetry breaking paradigm. The study of gapped domain walls (GDWs) of topological phases is crucial to understand the full nature of topological orders, and has potential applications in quantum computing and novel material properties. Although theoretical frameworks like low-energy effective topological quantum field theories (TQFTs) have offered comprehensive understanding for GDWs~\cite{kapustin2011topological,fuchs2013bicategories,lan2015gapped}, there remains a significant gap between these theories and their realization in actual physical systems. Nonetheless, lattice models of topological orders with GDWs bridge this gap by offering concrete, experimentally relevant models that allow for studying GDWs in greater detail~\cite{Kitaev-Kong-2012,jia2023boundary,zhao2023characteristic,christian2023lattice}.

In this paper, we develop a new way to build GDWs systematically in the framework of the Levin-Wen (LW) model. When we fold a topological system along its GDW, the GDW would become a gapped boundary (GB) of the folded system due to topological invariance. 
Given the complete classification of gapped boundaries (GBs) for 2+1D nonchiral topological phases in the Levin-Wen model through Frobenius algebras~\cite{hu2018boundary}---mathematically equivalent to the Kitaev-Kong formulation using module categories---we ask: Can we unfold a topological system with a GB and obtain a GDW between two (not necessarily different) topological phases? We tackle this question by considering two LW models, each having an open side specified by a certain subset of the model's input data, and sewing the two models along their open sides in a way compatible with the two subsets of input data, such that the resultant model is still gapped, exactly solvable, and topological. The sewed open sides will be a GDW, and different possible ways of sewing yield different types of GDWs.

As such, our key result is a lattice model that can describe all GDWs separating any two LW models with given two input unitary fusion categories (UFCs), therefore offering a unified framework of GDWs for 2+1D topological phases, including $e$-$m$ exchanging type of GDWs.
Additionally, our GDW construction stems from an unfolding process, as we will show that upon folding, a GDW would transform into a GB of the folded phase.

This paper is structured as follows. Section \ref{sec:Motivation} presents the motivation and intuition behind our construction. Section \ref{section: general constructions} introduces the LW model with GDWs, developing the definition of $\AAeta$-bimodules and showing that the Hamiltonians of GDWs are classified by the triples of input data $(A_1, A_2, \eta)$. Section \ref{section: excitation spectrum} constructs creation operators and corresponding measurement operators for the GDWs. Sections \ref{sec: GDWs in Toric code phase}, \ref{sec: GDWs in the doubled Ising}, and \ref{sec: GDWs in DI-TC} provide concrete examples of our construction. Finally, Section \ref{section: folding trick} applies the folding trick to our GDW construction, demonstrating its equivalence to an input Frobenius algebra that characterizes a GB of the folded lattice.

\section{Motivations and Sketch of Our Approach}\label{sec:Motivation}

A complete understanding of topological orders is impossible without understanding their GBs and GDWs, which also shed new light into topological quantum computation~\cite{beigi2011quantum,Cong-TQC-2017,cong2017universal}. The study of GBs and GDWs is essential for understanding anyon condensation, which exhibits intriguing physical consequences and complex mathematical structures~\cite{Bais2009,Bais2009-prl,Kong-2014,neupert2016boson,hu2022anyon,zhao2023characteristic,kesselring2024anyon,zhao2024symmetry,zhao2024nonabelian}. 
The lattice construction of GDWs in the LW model \cite{Kitaev-Kong-2012,jia2023boundary} as defect lines using bimodule categories requires extra input data on top of fusion categories and is difficult to construct interdomain and domain-wall ribbon operators. Furthermore, the recent approach to constructing GDWs through anyon condensation \cite{zhao2023characteristic} cannot describe $e$-$m$ exchanging GDWs \cite{bombin2010topological,buerschaper2013electric,wang2020electric,hu2020electric,huston2023composing}. Here, we take a different approach to construct GDWs in the LW model and offer a classification for GDWs, which includes anyon condensation induced GDWs and $e$-$m$ exchanging type of GDWs.

Gapped boundaries in the LW model have been completely classified by the Frobenius algebras of the model's input UFC $\mathcal{C}$~\cite{hu2018boundary}. Hence, a classification of GDWs in the LW model should comply with the classification of GBs in that when the system is folded along a GDW, a GDW would become a GB characterized by certain Frobenius algebra of the input UFC of the folded model. Conversely, suppose we can unfold an open LW model\footnote{An open LW model means a LW model with a GB.} along its GB, we would obtain a model with a GDW. The question is: How can we specify such a GDW? 

To tackle this question, we consider two LW models with respectively input UFCs $\mathcal{C}_1$ and $\mathcal{C}_2$, and each model has an open side (not necessarily a GB), see Fig. \ref{fig: gluing process}. We refer to the two models as the $\mathcal{C}_1$-model and the $\mathcal{C}_2$-model. It is natural to assume that the DOFs on the open side of either model are a subset of the bulk DOFs of the model. We thus specify the two open sides of the two models by algebra objects $A_1\in \mathcal{C}_1$ and $A_2\in \mathcal{C}_2$ for the following reason. We want to make our model as general as possible, so minimial assumptions should be imposed on the data specifying the open sides. It is then natural to choose algebra objects---the simplest yet interesting structures---in the input UFCs. Now we try to glue the two models along their open sides as follows. We device a joining function $\eta: A_1\times A_2 \to \mathbb{C}$, which dictates how the DOFs taking value in $A_1$ are coupled with those taking value in $A_2$. A compatibility check of this gluing is that when the $\mathcal{C}_2$-model ($\mathcal{C}_1$-model) describes a trivial phase, $A_1$ ($A_2$) should be a Frobenius algebra object in $\mathcal{C}_1$ ($\mathcal{C}_2$) and characterizes the GB of the $\mathcal{C}_1$-model ($\mathcal{C}_2$-model).

\begin{figure}[htbp]
    \centering
    \input{images/notation/fig__gluing_process}
    \caption{The gluing process. (a) Part of the lattice of the $\mathcal{C}_1$-model and that of the $\mathcal{C}_2$-model. Their open sides are characterized by $A_1$ and $A_2$, repectively. (b) A GDW constructed by gluing the two lattice along their open sides via the joining function $\eta$. }
    \label{fig: gluing process}
\end{figure}

Now, to ensure that the gluing process of two open sides results in a GDW, we should construct a proper Hamiltonian to describe the glued sides. We take the tailed LW model defined in \cite{hu2018full} because it has an enlarged Hilbert space that encompasses the full anyon spectra of the topological phases it describes. Based on this model, we construct the Hamiltonian of the glued sides, such that it is exactly solvable and gapped. This construction unravels the function $\eta$ that joins the algerbas $A_1$ and $A_2$ characterizing the two open sides. A tuple $(A_1,A_2,\eta)$ uniquely specifies a GDW.

As shown in Section \ref{section: general constructions}, upon joining two open edges, new DOFs inevitably emerge at the joining point. Such new DOFs form generalized bimodules of $A_1$ and $A_2$---dubbed $\AAeta$-bimodules---to be defined in Section \ref{Subsec: A1A2bimodule}. Given a GDW specified by $(A_1,A_2,\eta)$, quasiparticles on the GDW are in one-to-one correspondence with $\AAeta$-bimodules.

We also perform a consistency check of our construction as follows. Suppose we have constructed a model comprising a $\mathcal{C}_1$-model and a $\mathcal{C}_2$ model sandwiching a GDW specified by $(A_1,A_2,\eta)$. We show that folding this model along its GDW results in a $\mathcal{C}_1^\text{op} \boxtimes \mathcal{C}_2$-model with a GB specified by the Frobenius algebra object $A_1^\text{op}\times_{\eta}A_2$ in $\mathcal{C}_1^\text{op} \boxtimes \mathcal{C}_2$. Therefore, our construction of GDWs in the LW model can be regarded as a way of unfolding the $\mathcal{C}_1^\text{op} \boxtimes \mathcal{C}_2$-model with a GB along the GB. This may inspire attempts in unfolding a non-chiral topological phase along its GB to realize a lattice model of chiral topological orders. 

\section{Exactly solvable LW model with GDWs}\label{section: general constructions}
\begin{figure}[htbp]
    \centering
    \input{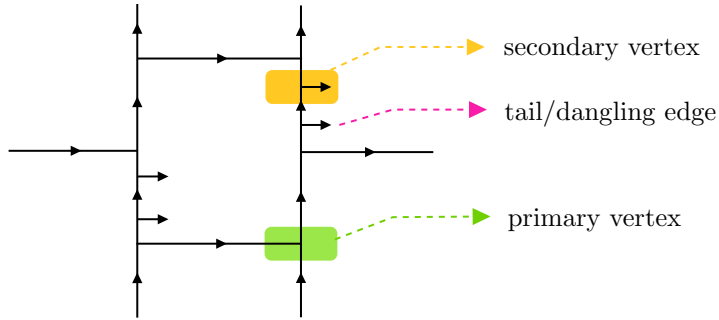}
    \caption{The tailed LW model. The primary and secondary vertices are highlighted in the lattice. }
    \label{fig: tailed_LW_lattice}
\end{figure}

The tailed Levin-Wen (LW) model \cite{hu2018full} is defined on a two-dimensional trivalent lattice (see Fig.~\ref{fig: tailed_LW_lattice}) with oriented edges and tails (dangling edges). The lattice consists of two types of vertices: \textbf{primary vertices}, which are trivalent with three incident edges, and \textbf{secondary vertices}, which have two incident edges and one tail. Each tail is associated with its nearest primary vertex.
Residing on each edge/tail is a DOF taking value in a finite set $L$ of labels that label the representative simple objects in a fusion category $\mathcal{C}$---the input fusion category of the model. We refer to this construction as the $\mathcal{C}$-model. The Hilbert space of the model is spanned by all possible label assignments to the edges and tails, subject to the constraint that the three labels meeting at any vertex must satisfy the fusion rules of $\mathcal{C}$.

The Hamiltonian of the model is the sum over all primary-vertex operators and plaquette operators. 
\begin{equation}
    H=-\sum_v A_v -\sum_p B_p. 
    \label{eq: original Hamiltonian}
\end{equation}
Detailed explanation of the Hamiltonian can be found in Appendix \ref{Appendix: the extended LW}. Here is a brief account. Each primary-vertex operator $A_v$ is a projector that acts on a primary vertex $v$. The action of $A_v$ returns $1$ if the label on the tail associated with $v$ is $1$---the trivial object of $\mathcal{C}$---otherwise, it vanishes. The plaquette operator $B_p$ is also a projector. All operators $A_v$ and $B_p$ commute with one another, such that the Hamiltonian is exactly solvable. A ground state of the system is a common eigenstate of all $A_v$ and $B_p$ with eigenvalue $+1$. In an excited state, however, the action of one or more $A_v$ and/or $B_p$ operators would return 0. As mentioned already, if the action of an $A_v$ returns 0, a chargeon is said to reside on the tail associated with $v$. If the action of a $B_p$ returns 0, a fluxon resides in plaquette $p$. If both the actions of a $B_p$ and an $A_v$ (with $v$'s tail in $p$) return $0$, a dyon inhabits in plaquette $p$.

We now cut the $\mathcal{C}$-model open vertically along edges and remove the part of the lattice to the right of and including the edges on the cut. This results in a model with an open side consisting of open edges (which should not be confused with tails in the bulk). The question arises: What are the DOFs on the open edges? It is known that if the open side is a GB, the DOFs on the open edges take values in a Frobenius algebra object in $\mathcal{C}$. This is referred to as the GB condition\cite{hu2018boundary,Hung-Wan-ADE-2015} or the Frobenius condition\cite{kock2004frobenius}. But our open side is not necessarily a GB, so it is reasonable to relax the Frobenius condition. We then assume that the DOFs on our open edges take value in an algebra object $A$ in $\mathcal{C}$ defined as a pair $A=(L_A,f)$, where $L_A\subseteq L$, and $f:L_A^3 \to \mathbb{C}$ is the algebra multiplication satisfying the defining properties:
\begin{equation}
\begin{aligned}
    \text{unit} \quad f_{b b^* 1}&=f_{b 1 b^*}=f_{1 b b^*}=1, \forall b\in L_{A} \\
    \text{cyclic} \quad f_{abc}&=f_{cab} . 
\end{aligned}
        \label{eq: properties of input algebra}
\end{equation}
We assume algebra $A$ is multiplicity free. Each open edge carries an element of $L_A$.

Next, given two fusion categories $\mathcal{C}_1$ and $\mathcal{C}_2$, with respectively label sets $L_1$ and $L_2$, consider the $\mathcal{C}_1$-model and $\mathcal{C}_2$, each having an open side. See Fig. \ref{fig: gluing process}(a). The open side of the $\mathcal{C}_1$-model ($\mathcal{C}_2$-model) is specified by the algebra objects $(L_{A_1},f)$ ($(L_{A_2},g)$), with $f$ ($g$) being the algebra multiplication defined by Eq. \eqref{eq: properties of input algebra}.

We are ready to glue the $\mathcal{C}_1$-model and $\mathcal{C}_2$-model along their open sides by joining the open edges head to head horizontally as seen in Fig. \ref{fig: gluing process}(b). We device a joining function $\eta: A_1\times A_2 \to \mathbb{C}$ that dictates joining two open edges: If $\eta_{ab} \neq 0$ for $a \in L_{A_1}$ and $b \in L_{A_2}$, the two open edges can be joined; otherwise, the joining is disallowed.

The total Hilbert space of the glued model contains only those states on the glued lattice with all open edges joined pairwise. See Fig. \ref{fig: gluing process}(b) for illustration. The glued lattice now has three regions: a vertical strip (grey in the figure) of plaquettes in the middle, which is to be proven as a GDW of the glued model, the original bulk of the $\mathcal{C}_1$-model on the left (red in the figure), and that of the $\mathcal{C}_2$-model on the right (blue in the figure). We will provide consistency conditions that determine all possible joining functions for given algebras $A_1$ and $A_2$. Each tuple $(A_1,A_2, \eta)$ specifies a GDW in the glued model. Hence, a glued model's input data is a 5-tuple $(\mathcal{C}_1, \mathcal{C}_2, A_1, A_2, \eta)$.

At this point, for a glued model with input data $(\mathcal{C}_1, \mathcal{C}_2, A_1, A_2, \eta)$, its total Hilbert space is spanned by configurations of all possible assignments of labels to the bulk edges (in black) on both sides of the GDW, pertaining to the fusion rules, and labels to the joined open edges (red line segments for the $\mathcal{C}_1$-model and blue for the $\mathcal{C}_2$-model) pertaining to the joining function $\eta$, which is symbolized by a square box in the figure \ref{fig: gluing process}(b). Nevertheless, this Hilbert space is insufficient to encompass quasiparticle excitations on the GDW, which have been shown to exist on the GDW between two topological phases, as well as on the GB of a topological phase\cite{hu2018full}. This insufficiency is due to that the DOFs on any two joined open edges are fully determined by the input joining function $\eta$. We thus need to enlarge the Hilbert space by introducing new DOFs to the GDW in the glued model to describe possible quasiparticle excitations on the GDW. But how? 

A clue can be drawn from the case of the LW model with a GB, which is characterized by a Frobenius algebra object of the model's input fusion category. In this case, the quasiparticles on the GB carry the bimodules of the Frobenius algebra. Coming back to our construction, it is then reasonable to guess that the quasiparticles on a GDW specified by $(A_1,A_2,\eta)$ carry certain bimodules acted on by both $A_1$ and $A_2$ in a way pertaining to $\eta$. Indeed, by constructing the quasiparticle creation operators (see Section \ref{section: excitation spectrum}) on our GDW, we can show that GDW quasiparticles do carry what we call $\AAeta$-bimodules, a generalized type of algebra bimodules. This finding motivates us to introduce new DOFs capturing such bimodules in the glued region of our lattice, such that the model's Hilbert space can encompass GDW quasiparticles in the first place.

To see how new DOFs would be introduced in the GDW, we first need to define $\AAeta$-bimodules as follows.

\subsection{Definition of \texorpdfstring{$\AAeta$}{}-bimodules \label{Subsec: A1A2bimodule}}

Consider two UFCs $\mathcal{C}_1, \mathcal{C}_2$ and two algebras $A_1 \in \mathcal{C}_1, A_2\in \mathcal{C}_2$. Let the set of all representative simple objects of $\mathcal{C}_1$, $\mathcal{C}_2$, $A_1$, $A_2$ be $L_1$, $L_2$, $L_{A_1}$, $L_{A_2}$, respectively. 
Given a joining function $\eta: A_1\times A_2 \rightarrow \mathbb{C}$, we can define a set
\begin{equation}
    L_{\eta}=\{(a,b)|\eta_{ab}\neq 0, a\in L_{A_1}, b\in L_{A_2}\}
    \label{eq: L_eta}
\end{equation}
to label the pairs of objects in $L_{A_1}$ and $L_{A_2}$ that are allowed by $\eta$. 
For convenience, we set a characteristic function $\Delta:\mathbb{C}\to\{0,1\}$ for $\eta$: 
\begin{equation}
    \Delta(x)=
    \begin{cases}
0,& x=0;\\
1,& x\neq 0.\\
    \end{cases}
    \label{eq: Delta function}
\end{equation}

We shall consider, without loss of generality, two scenarios: (1) $\mathcal{C}_2$ is a subcategory of $\mathcal{C}_1$, then $L_2 \subset L_1$ and all other data of $\mathcal{C}_2$ are embedded in those of $\mathcal{C}_1$. (2) $\mathcal{C}_2$ is not a subcategory of $\mathcal{C}_1$, but the objects and morphisms of $\mathcal{C}_2$ can be represented by those of $\mathcal{C}_1$. For instance, when $\mathcal{C}_1$ is Morita equivalent to $\mathcal{C}_2$ (e.g. $\mathcal{R}ep(\mathbb{Z}_1)$ and $\mathcal{V}ec(\mathbb{Z}_2)$), one can establish isomorphic maps between their input data. Then, we define $\AAeta$-bimodules as follows.

An $\AAeta$-bimodule $M$ is a pair $(L_M, P_M)$, where $L_M\subseteq L_1$, and $P_M$ is a set of action tensors $[P_M]^{ab}_{i m j}$ labeled by $(a,b)\in L_\eta$ and tensorial indices $i,j\in L_M, m \in L_1$, satisfying the following equation:
    \vspace{-0.5em}
\begin{equation}
    \input{images/DW_excitations/fig_definition_of_A1-eta-A2-bimodules_1} = \Delta(\eta_{a_1 b_1} \eta_{a_2 b_2} \eta_{a_3 b_3}) \input{images/DW_excitations/fig_definition_of_A1-eta-A2-bimodules_2}.
    \label{eq: definiton of A1-A2-etabimodule}
\end{equation}    \vspace{-1em}

In Eq. \eqref{eq: definiton of A1-A2-etabimodule} and hereafter, the box labeled by $P_M$ encapsulates the corresponding action tensor:
\begin{equation}
    \input{images/DW_excitations/fig__P-graphical_notation_1} = \sum_{m \in L_1} [P_M]_{s_1 m s_2}^{ab}  \input{images/DW_excitations/fig__P-graphical_notation_2}, \label{eq: linear combination of P tensors}
\end{equation}
where $s_1$, $s_2\in L_M$, and $m\in L_1$. 

In Eq. \eqref{eq: definiton of A1-A2-etabimodule} and hereafter, a thickened edge indicates a summation over all values of this edge's DOF, while a thick red (blue) dot at a vertex absorbs the algebra multiplication $f$ ($g$) for $A_1$ ($A_2$): 
\begin{equation}
   \tikzset{every picture/.style={line width=0.75pt}} 
\begin{tikzpicture}[x=0.75pt,y=0.75pt,yscale=-1,xscale=1, baseline=(XXXX.south) ]
\path (0,57);\path (75.31640625,0);\draw    ($(current bounding box.center)+(0,0.3em)$) node [anchor=south] (XXXX) {};
\draw [color={rgb, 255:red, 208; green, 2; blue, 27 }  ,draw opacity=1 ][line width=0.75]    (56.75,52.52) -- (34.71,30.48) ;
\draw [shift={(47.78,43.55)}, rotate = 225] [fill={rgb, 255:red, 208; green, 2; blue, 27 }  ,fill opacity=1 ][line width=0.08]  [draw opacity=0] (5.36,-2.57) -- (0,0) -- (5.36,2.57) -- (3.56,0) -- cycle    ;
\draw [color={rgb, 255:red, 208; green, 2; blue, 27 }  ,draw opacity=1 ][line width=0.75]    (12.74,52.45) -- (34.71,30.48) ;
\draw [shift={(21.67,43.52)}, rotate = 315] [fill={rgb, 255:red, 208; green, 2; blue, 27 }  ,fill opacity=1 ][line width=0.08]  [draw opacity=0] (5.36,-2.57) -- (0,0) -- (5.36,2.57) -- (3.56,0) -- cycle    ;
\draw [color={rgb, 255:red, 208; green, 2; blue, 27 }  ,draw opacity=1 ][line width=0.75]    (34.71,3.71) -- (34.71,30.48) ;
\draw [shift={(34.71,14.19)}, rotate = 90] [fill={rgb, 255:red, 208; green, 2; blue, 27 }  ,fill opacity=1 ][line width=0.08]  [draw opacity=0] (5.36,-2.57) -- (0,0) -- (5.36,2.57) -- (3.56,0) -- cycle    ;
\draw  [color={rgb, 255:red, 208; green, 2; blue, 27 }  ,draw opacity=1 ][fill={rgb, 255:red, 208; green, 2; blue, 27 }  ,fill opacity=1 ] (31.75,30.48) .. controls (31.75,28.85) and (33.08,27.53) .. (34.71,27.53) .. controls (36.34,27.53) and (37.67,28.85) .. (37.67,30.48) .. controls (37.67,32.12) and (36.34,33.44) .. (34.71,33.44) .. controls (33.08,33.44) and (31.75,32.12) .. (31.75,30.48) -- cycle ;
\draw (54.28,35.51) node [anchor=north west][inner sep=0.75pt]  [font=\small]  {$b$};
\draw (8.91,35.36) node [anchor=north west][inner sep=0.75pt]  [font=\small]  {$a$};
\draw (38.71,5.41) node [anchor=north west][inner sep=0.75pt]  [font=\small]  {$c$};
\end{tikzpicture}
 = f_{abc} \tikzset{every picture/.style={line width=0.75pt}} 
\begin{tikzpicture}[x=0.75pt,y=0.75pt,yscale=-1,xscale=1, baseline=(XXXX.south) ]
\path (0,61);\path (75.31640625,0);\draw    ($(current bounding box.center)+(0,0.3em)$) node [anchor=south] (XXXX) {};
\draw [color={rgb, 255:red, 208; green, 2; blue, 27 }  ,draw opacity=1 ][line width=0.75]    (58.08,53.19) -- (36.04,31.15) ;
\draw [shift={(49.11,44.22)}, rotate = 225] [fill={rgb, 255:red, 208; green, 2; blue, 27 }  ,fill opacity=1 ][line width=0.08]  [draw opacity=0] (5.36,-2.57) -- (0,0) -- (5.36,2.57) -- (3.56,0) -- cycle    ;
\draw [color={rgb, 255:red, 208; green, 2; blue, 27 }  ,draw opacity=1 ][line width=0.75]    (14.07,53.12) -- (36.04,31.15) ;
\draw [shift={(23.01,44.18)}, rotate = 315] [fill={rgb, 255:red, 208; green, 2; blue, 27 }  ,fill opacity=1 ][line width=0.08]  [draw opacity=0] (5.36,-2.57) -- (0,0) -- (5.36,2.57) -- (3.56,0) -- cycle    ;
\draw [color={rgb, 255:red, 208; green, 2; blue, 27 }  ,draw opacity=1 ][line width=0.75]    (36.04,4.37) -- (36.04,31.15) ;
\draw [shift={(36.04,14.86)}, rotate = 90] [fill={rgb, 255:red, 208; green, 2; blue, 27 }  ,fill opacity=1 ][line width=0.08]  [draw opacity=0] (5.36,-2.57) -- (0,0) -- (5.36,2.57) -- (3.56,0) -- cycle    ;
\draw (55.61,36.17) node [anchor=north west][inner sep=0.75pt]  [font=\small]  {$b$};
\draw (10.24,36.03) node [anchor=north west][inner sep=0.75pt]  [font=\small]  {$a$};
\draw (40.04,6.07) node [anchor=north west][inner sep=0.75pt]  [font=\small]  {$c$};
\end{tikzpicture}
, \quad 
    \tikzset{every picture/.style={line width=0.75pt}} 
\begin{tikzpicture}[x=0.75pt,y=0.75pt,yscale=-1,xscale=1, baseline=(XXXX.south) ]
\path (0,59);\path (75.98307291666671,0);\draw    ($(current bounding box.center)+(0,0.3em)$) node [anchor=south] (XXXX) {};
\draw [color={rgb, 255:red, 33; green, 55; blue, 191 }  ,draw opacity=1 ][line width=0.75]    (58.75,53.85) -- (36.71,31.82) ;
\draw [shift={(49.78,44.89)}, rotate = 225] [fill={rgb, 255:red, 33; green, 55; blue, 191 }  ,fill opacity=1 ][line width=0.08]  [draw opacity=0] (5.36,-2.57) -- (0,0) -- (5.36,2.57) -- (3.56,0) -- cycle    ;
\draw [color={rgb, 255:red, 33; green, 55; blue, 191 }  ,draw opacity=1 ][line width=0.75]    (14.74,53.79) -- (36.71,31.82) ;
\draw [shift={(23.67,44.85)}, rotate = 315] [fill={rgb, 255:red, 33; green, 55; blue, 191 }  ,fill opacity=1 ][line width=0.08]  [draw opacity=0] (5.36,-2.57) -- (0,0) -- (5.36,2.57) -- (3.56,0) -- cycle    ;
\draw [color={rgb, 255:red, 33; green, 55; blue, 191 }  ,draw opacity=1 ][line width=0.75]    (36.71,5.04) -- (36.71,31.82) ;
\draw [shift={(36.71,15.53)}, rotate = 90] [fill={rgb, 255:red, 33; green, 55; blue, 191 }  ,fill opacity=1 ][line width=0.08]  [draw opacity=0] (5.36,-2.57) -- (0,0) -- (5.36,2.57) -- (3.56,0) -- cycle    ;
\draw  [color={rgb, 255:red, 33; green, 55; blue, 191 }  ,draw opacity=1 ][fill={rgb, 255:red, 33; green, 55; blue, 191 }  ,fill opacity=1 ] (33.75,31.82) .. controls (33.75,30.18) and (35.08,28.86) .. (36.71,28.86) .. controls (38.34,28.86) and (39.67,30.18) .. (39.67,31.82) .. controls (39.67,33.45) and (38.34,34.77) .. (36.71,34.77) .. controls (35.08,34.77) and (33.75,33.45) .. (33.75,31.82) -- cycle ;
\draw (56.28,36.84) node [anchor=north west][inner sep=0.75pt]  [font=\small]  {$b$};
\draw (10.91,36.69) node [anchor=north west][inner sep=0.75pt]  [font=\small]  {$a$};
\draw (40.71,6.74) node [anchor=north west][inner sep=0.75pt]  [font=\small]  {$c$};
\end{tikzpicture}
 = g_{abc} \tikzset{every picture/.style={line width=0.75pt}} 
\begin{tikzpicture}[x=0.75pt,y=0.75pt,yscale=-1,xscale=1, baseline=(XXXX.south) ]
\path (0,58);\path (73.31640625,0);\draw    ($(current bounding box.center)+(0,0.3em)$) node [anchor=south] (XXXX) {};
\draw [color={rgb, 255:red, 33; green, 55; blue, 191 }  ,draw opacity=1 ][line width=0.75]    (56.75,51.85) -- (34.71,29.82) ;
\draw [shift={(47.78,42.89)}, rotate = 225] [fill={rgb, 255:red, 33; green, 55; blue, 191 }  ,fill opacity=1 ][line width=0.08]  [draw opacity=0] (5.36,-2.57) -- (0,0) -- (5.36,2.57) -- (3.56,0) -- cycle    ;
\draw [color={rgb, 255:red, 33; green, 55; blue, 191 }  ,draw opacity=1 ][line width=0.75]    (12.74,51.79) -- (34.71,29.82) ;
\draw [shift={(21.67,42.85)}, rotate = 315] [fill={rgb, 255:red, 33; green, 55; blue, 191 }  ,fill opacity=1 ][line width=0.08]  [draw opacity=0] (5.36,-2.57) -- (0,0) -- (5.36,2.57) -- (3.56,0) -- cycle    ;
\draw [color={rgb, 255:red, 33; green, 55; blue, 191 }  ,draw opacity=1 ][line width=0.75]    (34.71,3.04) -- (34.71,29.82) ;
\draw [shift={(34.71,13.53)}, rotate = 90] [fill={rgb, 255:red, 33; green, 55; blue, 191 }  ,fill opacity=1 ][line width=0.08]  [draw opacity=0] (5.36,-2.57) -- (0,0) -- (5.36,2.57) -- (3.56,0) -- cycle    ;
\draw (54.28,34.84) node [anchor=north west][inner sep=0.75pt]  [font=\small]  {$b$};
\draw (8.91,34.69) node [anchor=north west][inner sep=0.75pt]  [font=\small]  {$a$};
\draw (38.71,4.74) node [anchor=north west][inner sep=0.75pt]  [font=\small]  {$c$};
\end{tikzpicture}
. 
    \label{eq: notation-thick dots}
\end{equation}

The Eq. \eqref{eq: definiton of A1-A2-etabimodule} indicates the equivalence between $A_1 \otimes (A_1\otimes M \otimes A_2)\otimes A_2 $ and $ (A_1\otimes A_1) \otimes M \otimes (A_2\otimes A_2)$.
We can transform the basis on the LHS to that on the RHS by a series of F-moves in the following:
\begin{equation}
    \input{images/domain_wall/fig__pachner_move2to2-1}  = \sum_{n\in L} G^{jim}_{kln} \, v_m \, v_n  \input{images/domain_wall/fig__pachner_move2to2-2} ,
    \label{eq: pachner move 2to2-2}
\end{equation}
\begin{equation}
    \input{images/domain_wall/fig__pachner_move2to2-3}  = \sum_{a_3 \in L_{A_1}} G^{j a_1 m}_{a_2 l a_3} \, v_m \, v_{a_3}  \input{images/domain_wall/fig__pachner_move2to2-4} ,
    \label{eq: pachner move 2to2-4}
\end{equation}
\begin{equation}
    \input{images/domain_wall/fig__pachner_move2to2-5}  = \sum_{b_3 \in L_{A_2}} G^{b_1 i m}_{k b_2 b_3} \,
     v_m \, v_{b_3}  \input{images/domain_wall/fig__pachner_move2to2-6}.
    \label{eq: pachner move 2to2-6}
\end{equation}

Then, comparing the linear combination coefficients on both sides of Eq.~\eqref{eq: definiton of A1-A2-etabimodule} yields the following tensor equation:
\begin{multline}
       \sum_{\stackrel{a_3\in L_{A_1},b_3\in L_{A_2}}{s_2\in L_M,l,m,n\in L_1}} P^{a_1 b_1}_{s_1 m s_2} P^{a_2 b_2}_{s_2 n s_3} G_{b_2 n s_2}^{a_1 m l} G_{s_3 a_2 n}^{a_1 l a_3} G_{b_2 l m}^{s_1 b_1 b_3} v_l v_{m} v_{n} v_{s_2} v_{a_3} v_{b_3} \\ =\sum_{\stackrel{a_3\in L_{A_1},b_3\in L_{A_2}}{l\in L_1}} \Delta(\eta_{a_1 b_1} \eta_{a_2 b_2} \eta_{a_3 b_3}) P_{s_1 l s_3}^{a_3 b_3} f_{a_2 a_1 a_3} g_{b_1 b_2 b_3},
       \label{eq: definiton of A1-A2-etabimodule formula}
\end{multline}
for $a_1,a_2\in L_{A_1}$, $b_1,b_2\in L_{A_2}$.

An $\AAeta$-bimodule is, in general, not a $A_1$-$A_2$-bimodule as defined in \cite{hu2018boundary}, but a generalization thereof. Unlike $A_1$-$A_2$-bimodules, an $\AAeta$-bimodule is not necessarily both a left $A_1$-module and right $A_2$-module. Nonetheless, when $\Delta(\eta_{ab})=1$ for any $a\in A_1, b\in A_2$, Eq. \eqref{eq: definiton of A1-A2-etabimodule formula} reduces to the defining equation of $A_1$-$A_2$-bimodules. 

The $A_1 \overset{\eta}{-} A_2$-bimodules satisfy the following orthonormality condition:
\begin{equation}
    \mathcal{T} \Big( \input{images/domain_wall/eq-orthonormality_condition_for_P_1} \Big) = \delta_{M,N} \Delta(\eta_{a_1 b_1} \eta_{a_2 b_2} \eta_{a_3 b_3}) \frac{d_{A_1} d_{A_2} d_s}{d_{M}} \Big( \input{images/domain_wall/eq-orthonormality_condition_for_P_2} \Big), 
\end{equation}
where $d_M=\sum_{s\in L_M} d_s$ and $d_{A}=\sum_{a\in A} d_a$. 

We package the three black edges that describe the DOFs before and after the action of $A_1$ and $A_2$ into a filled box. As shown in Eq. \eqref{eq: filled square box}, the filled box associated with an arbitrary $\AAeta$-bimodule $M$ can be rigorously expanded as a linear combination of basis states, which explicitly capture the internal DOFs in the filled box.
\begin{equation}
        \Big(\input{images/DW_excitations/fig__inner_dof_of_box} \Big) = \sum_{s_1,s_2\in L_{M}} \frac{u_{s_1} u_{s_2}}{u_{a} u_{b}} \Big(\input{images/DW_excitations/fig__P-graphical_notation_1} \Big),
        \label{eq: filled square box}
\end{equation}
where $u_i=(d_i)^{\frac{1}{4}}$ and $d_i$ is the quantum dimension of the string type $i$. 
In cases where the trivial $\AAeta$-bimodule $M_0$ is explicitly required, we use an unfilled square box to avoid ambiguity.

The trivial $\AAeta$-bimodule $M_0$ is a pair $(L_{M_0},P_{M_0})$, where $L_{M_0}$ contains $1 \in L_1$, and the action tensors satisfying
\begin{equation}
    [P_{M_0}]_{1m1}^{ab} = \delta_{a,m} \delta_{b,m},
\end{equation}
which indicates that the trivial anyon in the bulk would be identified as $M_0$ in the GDW.

\subsection{Hilbert Space and Hamiltonian} \label{subsection: Pachner moves at GDW}

\begin{figure}[htbp]
    \centering
    \input{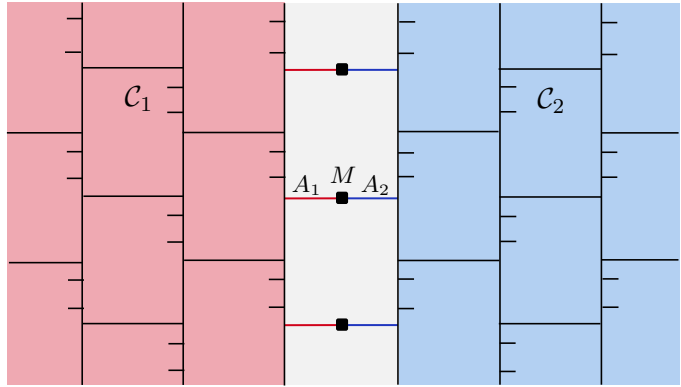}
    \caption{Two extended LW models (red and blue) with input fusion categories $\mathcal{C}_1$ and $\mathcal{C}_2$, joint by a DW (light grey). The joining points, bearing the algebraic actions of $A_1$ and $A_2$, are characterized by the $\AAeta$-bimodules $M$. }
    \label{fig: Hilbert space}
\end{figure}

To enlarge the Hilbert space, such that it can bear quasiparticle exitations along the DW, we dress each gluing point by a filled square box as a new DOF, taking value in the simple (i.e., irreducible) $\AAeta$ bimodules, as shown in Fig. \ref{fig: Hilbert space}. The enlarged Hilbert space is then spanned by all possible configurations (subject to fusion rules at all vertices) of the simple objects of the input UFCs $\mathcal{C}_1$ and $\mathcal{C}_2$ on the bulk edges, the basis elements of the algebras $A_1$ and $A_2$ on the DW edges, and the simple $\AAeta$ bimodules $\{M_0, M_1, M_2, \dots\}$ on the filled square boxes.

For the tailed LW model with the DW shown in Fig.~\ref{fig: Hilbert space}, the modified Hamiltonian from \eqref{eq: original Hamiltonian} is given by:
\begin{equation}
    H=-\sum_{v\text{ in bulk}} A_v^\text{bulk} -\sum_{p\text{ in bulk}} B_p^\text{bulk} -\sum_{v\text{ in DW}} A_v^\text{DW}  -\sum_{p\text{ in DW}} B_p^\text{DW}. 
    \label{eq: Hamiltonian with DW}
\end{equation}
The operators involved in the Hamiltonian are explained as follows. Any primary vertex with black incident edges is a bulk vertex, and one with a colored incident edge is a DW vertex. Bulk and DW plaquettes are obvious. The actions of the bulk operators $A_v^\text{bulk}$ and $B_p^\text{bulk}$ of the $\mathcal{C}_1$-model and the $\mathcal{C}_2$-model are given in Appendix~\ref{Appendix: the extended LW}.

Then the vertex operators for the vertices in the GDW are given as follows:
\begin{equation}
    A_v^\text{DW} \ket[Big]{\input{images/domain_wall/fig-Av_op_in_GDW_1}}=\delta_{q,0} \ket[Big]{\input{images/domain_wall/fig-Av_op_in_GDW_1}} ,\quad   A_v^\text{DW} \ket[Big]{\input{images/domain_wall/fig-Av_op_in_GDW_2}}=\delta_{q,0} \ket[Big]{\input{images/domain_wall/fig-Av_op_in_GDW_2}}. 
    \label{eq: vertex operators in GDW}
\end{equation}
Plaquette operators on the GDW is defined as
\begin{equation}
B_p^\text{DW}=\sum_{s\in A_1, t\in A_2} v_s v_t (\eta_{st})^2 B_p^{st},    \label{eq: DW plaquette op}
\end{equation}
where $v_i=\sqrt{d_i}$ with $d_i$ the quantum dimension of $i$. An operator $B_p^{st}$ acts on a DW plaquette by inserting a bicolored loop in plaquette $p$ and fusing the loop with $p$'s boundary edges. This action is given as follows. 
\begin{equation}
\begin{aligned}
    B_p^{st}  \Bigg| \input{images/domain_wall/fig-Domain_wall_Bp_1} \Vast\rangle = \delta_{M_0, M_1} \Delta(\eta_{i_1 j_1}\eta_{i_4 j_4}) \, \Bigg| \input{images/domain_wall/fig-Domain_wall_Bp_2} \Vast\rangle & \\  =  \sum_{\substack{i_1^\prime,i_2^\prime,i_3^\prime,i_4^\prime\in A_1 \\ j_1^\prime,j_2^\prime,j_3^\prime,j_4^\prime\in A_2}} 
    \Delta(\eta_{i_1 j_1} \eta_{i_4,j_4} \eta_{i_1^\prime i_1^\prime} \eta_{j_4^\prime j_4^\prime})
    u_{i_1} u_{j_1} u_{i_4} u_{j_4} u_{i_1^\prime} u_{j_1^\prime} 
    u_{i_4^\prime} u_{j_4^\prime} v_{i_2} v_{j_2} & \\ \times   v_{i_3} v_{j_3} v_{i_2^\prime} v_{j_2^\prime} v_{i_3^\prime} v_{j_3^\prime} 
    G^{s^* {i_1^\prime}^* i_1}_{k_1^* i_2 {i_2^\prime}^*} G^{s^*{i_2^\prime}^* i_2}_{k_2 i_3 {i_3^\prime}^*} G^{s^* {i_3^\prime}^* i_3}_{k_3 i_4 {i_4^\prime}^*} G^{t^* j_4^\prime j_4^*}_{l_3 j_3^* j_3^\prime} G^{t^* j_3^\prime j_3^*}_{l_2^* j_2^* j_2^\prime} & \\ \times  G^{t^* j_2^\prime j_2^*}_{l_1^* j_1 {j_1^\prime}^*} 
    f_{s i_1^* i_1^\prime} f_{s^* i_4^\prime i_4^*} g_{t^* {j_1^\prime}^* j_1} g_{t j_4 {j_4^\prime}^*} \Bigg| \input{images/domain_wall/fig-Domain_wall_Bp_3} \Vast\rangle & . 
\end{aligned}
\label{eq: Bp^st operator}
\end{equation} 
In the derivation above, as a simplification, we moved any tail attached to $p$'s boundary edges away using F-moves Eqs.~\eqref{eq: pachner move 2to2-2}--\eqref{eq: pachner move 2to2-6}, such that the computation involves no tails.
Equation \eqref{eq: Bp^st operator} implies that DW plaquette operators project out states where filled square boxes in the DW carry non-trivial $\AAeta$-bimodules.

The derivation of the plaquette operators needs the Pachner moves in the GDW, which has not been defined yet. Here, we propose the Pacher moves at the GDW as follows:
\begin{align}
        & \mathcal{T}\ket[\Big]{\input{images/Review/fig__DW_pachner_move_1}}= \Delta[\eta_{j^\prime j} \eta_{i^\prime i} \eta_{k^\prime k}] \frac{u_{i^\prime}u_{j^\prime}u_i u_j}{u_k u_{k^\prime}} f_{{i^\prime} {j^\prime} {k^\prime}^*} g_{j^* i^* k} \ket[\Big]{\input{images/Review/fig__DW_pachner_move_2}},  \label{eq: DW Pachner move} \\
        & \mathcal{T} \ket[\Big]{\input{images/Review/fig__DW_pachner_move_3}} = \sum_{\substack{i^\prime, j^\prime \in A_1 \\ i, j \in A_2}} \Delta[\eta_{j^\prime j} \eta_{i^\prime i} \eta_{k^\prime k}] \frac{u_{i^\prime}u_{j^\prime}u_i u_j}{u_k u_{k^\prime}} f_{{i^\prime} {j^\prime} {k^\prime}^*} g_{j^* i^* k} \ket[\Big]{\input{images/Review/fig__DW_pachner_move_4}}. \label{eq: DW Pachner move-reverse}
\end{align}
Equation \eqref{eq: DW Pachner move} is equivalent to the defining equation \eqref{eq: definiton of A1-A2-etabimodule} of $\AAeta$-bimodules. 

We have not yet discussed which joining functions are appropriate to guarantee compatibility between the DW and the adjancent $\mathcal{C}_1$-model and $\mathcal{C}_2$-model. We now tackle this problem to derive all possible joining functions.

\subsection{Exact solvability and condition on joining functions 
\label{section: Exact solvability}
}

In order that the model is exactly solvable, we require the operators in the DW to commute among themselves and with bulk operators. Domain-wall operators by definition commute with all bulk operators. Domain-wall vertex operators also commute with themselves and with DW plaquette operators. The commutativity between two neighboring DW plaquette operators is nontrivial and leads to the following condition:
\begin{multline}
    \sum_{\substack{k_2,k_2^{\prime}\in A_1 \\ l_2,l_2^{\prime}\in A_2}}\sum_{\substack{\eta_{i_1 j_1},\eta_{i_2 j_2},\eta_{k_1 l_1},\\ \eta_{k_2 l_2},\eta_{k_3 l_3} \neq 0} } G^{k_3 i_2 k_2^*}_{k_1^* i_1^* {k_2^{\prime}}^*} G^{j_1 l_1 l_2^*}_{j_2^* l_3^* l_2^{\prime}} v_{l_2} v_{l_2^{\prime}} v_{k_2} v_{k_2^{\prime}} f_{k_2^*k_3i_2}f_{k_1^*i_1^*k_2}g_{l_2^*j_1l_1}g_{l_3^*l_2j_2^*} \\ =\sum_{\substack{k_2^\prime \in A_1 \\ l_2^{\prime}\in A_2}}\sum_{\substack{\eta_{i_1 j_1},\eta_{i_2 j_2},\eta_{k_1 l_1},\\ \eta_{k_2^\prime l_2^\prime}, \eta_{k_3 l_3} \neq 0} } f_{{k_2^{\prime}}^*i_1^*k_3} f_{k_1^*k_2^\prime i_2} g_{{l_2^{\prime}}^*l_1j_2^*} g_{l_3^*j_1l_2^{\prime}},
    \label{eq: domain wall Bp commutativity}    
\end{multline}
which can be presented graphically as: 
\begin{equation}
    \mathcal{T} \ket[\Big]{\input{images/domain_wall/fig-Bp_commutativity_condition_1}} = \Delta[\eta_{i_1 j_1} \eta_{i_2 j_2} \eta_{k_1 l_1} \eta_{k_3 l_3}] \ket[\Big]{\input{images/domain_wall/fig-Bp_commutativity_condition_2}},
    \label{eq: domain wall Bp commutativity-graphical} 
\end{equation}

Vertex operators $A_v$ and plaquette operators $B_p$ in the original LW model are projectors, which detect whether an anyon excitation exists where such an operator acts. As we now extend the LW model to the case with GDW, we also require all DW operators to be projectors. Domain-wall vertex operators $A_v^\text{DW}$ are projectors by definition. Demanding DW plaquette operators $B_p^\text{DW}$ to be projectors results in the following condition on the joining function $\eta$:
\begin{equation}
    \sum_{\substack{i^\prime ,j^\prime \in A_1 \\i,j\in A_2}} \sum_{\substack{\eta_{i^\prime i},\eta_{j^\prime j},\\ \eta_{k^\prime k}\neq 0}} \delta_{i^\prime j^\prime k^\prime }\delta_{ijk}\frac{v_{i^\prime }v_{j^\prime } v_i v_j}{v_k v_{k^\prime }}f_{{i^\prime }^*{j^\prime }^*k^\prime }f_{j^\prime i^\prime {k^\prime }^*}g_{jik^*}g_{i^*j^*k} (\eta_{j^\prime j}\eta_{i^\prime i})^2
    = (\eta_{k^\prime k})^2,
    \label{eq: domain wall Bp projective}
\end{equation}

See Appendix~\ref{appendix: exactly solvable conditions for Bp} for the derivation of \eqref{eq: domain wall Bp commutativity} and \eqref{eq: domain wall Bp projective}. Appendix \ref{Appendix: specific solutions for eta} shows the solutions of the joining functions $\eta$ in certain cases. We find that even when $\mathcal{C}_1$ and $\mathcal{C}_2$ are fundamentally different, there still exist solutions for $\eta$ that physically connect the GBs of the $\mathcal{C}_1$-model and the $\mathcal{C}_2$-model. In this case, $A_1$ and $A_2$ should be Frobenius algebra objects in $\mathcal{C}_1$ and $\mathcal{C}_2$, respectively, indicating that the GDW is the vacuum.

Now that our Hamiltonian \eqref{eq: Hamiltonian with DW} consists of commuting projectors, the system is gapped. The ground-state Hilbert space is topologically protected, in the sense that it is invariant under the Pachner moves. Hence, our Hamiltonian \eqref{eq: Hamiltonian with DW} defines a composite system of two LW models separated by a GDW. A GDW between two LW models with input UFCs $\mathcal{C}_1$ and $\mathcal{C}_2$ is specified by a triple $(A_1,A_2,\eta)$, where $A_1$ ($A_2$) is an algebra object of $\mathcal{C}_1$ ($\mathcal{C}_2$), and $\eta$ is a pairing function of the two algberas. 

\section{Ground states and GDW Excitations \label{section: excitation spectrum}}

The ground states of the tailed LW model with GDWs are the $+1$ eigenvectors of all vertex operators and plaquette operators in the Hamiltonian \eqref{eq: Hamiltonian with DW}. Therefore, in a ground state, the tails in Fig. \ref{fig: Hilbert space} are all trivial and the boxes at the gluing points are all labeled by the trivial $\AAeta$-bimodule $M_0$. For the ground states, an edge along the left (right) boundary of the GDW region becomes a right $A_1$-module in $\mathcal{C}_1$ (left $A_2$-module in $\mathcal{C}_2$). A right $A_1$-module (left $A_2$-module) in $\mathcal{C}_1$ ($\mathcal{C}_2$) is a subset $L_{\mathrm{RMod}_{A_1}(\mathcal{C}_1)} \subset L$ ($L_{\mathrm{LMod}_{A_2}(\mathcal{C}_2)} \subset L$ ) equipped with a right (left) action tensor $[\rho_1]_{i_1 i_2}^{a_1}$ ($[\rho_2]_{j_1 j_2}^{b_1}$) satisfying the following equations:
\begin{equation}
    \mathcal{T} \ket[\Big]{\input{images/DW_excitations/fig_A1_module_1}}=\ket[\Big]{\input{images/DW_excitations/fig_A1_module_2}}, \quad \mathcal{T} \ket[\Big]{\input{images/DW_excitations/fig_A2_module_1}}=\ket[\Big]{\input{images/DW_excitations/fig_A2_module_2}}.
    \label{eq: A-module}
\end{equation}

In the tailed LW model, the elementary excitations (dyon species) are endowed with topological symmetries\footnote{The elementary excitations are identified with irreducible representations of the tube algebra \cite{hu2018full}.}, which can be characterized by irreducible solutions of half-braiding tensors in Appendix~\ref{Appendix: ribbon operators in the bulk}.
At the GDW, elementary excitations are characterized by irreducible $A_1 \overset{\eta}{-} A_2$-bimodules. GDW quasiparticles can be pairwise created by the following creation operator: 
\begin{equation}
\begin{aligned}
        W_{M}^{\text{gp}_1} & \Bigg| \input{images/DW_excitations/fig__creation_operators_at_the_domain_wall_1} \Vast\rangle = \Bigg| \input{images/DW_excitations/fig__creation_operators_at_the_domain_wall_2} \Vast\rangle.
\end{aligned}
\label{eq: DW creation operator}
\end{equation}
Here, the state on the LHS with all $M_0$ at the GDW indicates the absence of quasiparticle excitations at the GDW. The action of $W_{M}^{\text{gp}_1}$ replaces the unfilled square box at the gluing point $\text{gp}_1$ by a filled square box associated with an irreducible $\AAeta$-bimodule $M$. As a result, a pair of quasiparticles of the same type, characterizing by the $\AAeta$-bimodule $M$, are created in the neighboring plaquettes $p_1$ and $p_2$ sharing the gluing point gp1 in the GDW. 

To detect these quasiparticles, we introduce the measurement operator $\Pi_M^{p_1}$, which acts as a projector on the excited state with a quasiparticle of type $M$ in the GDW plaquette $p_1$. More detailed derivation is given in Appendix~\ref{appendix: Measuring operator at the domain wall}.
\begin{equation}
        \Pi_{M}^{p_1} \Bigg| \input{images/DW_excitations/fig__measuring_operator_A1-A2-bimodule_1} \Vast\rangle = \Bigg| \input{images/DW_excitations/fig__measuring_operator_A1-A2-bimodule_2} \Vast\rangle = \delta_{M,M^\prime} \Bigg| \input{images/DW_excitations/fig__measuring_operator_A1-A2-bimodule_3} \Vast\rangle.
\end{equation}

In the next sections, we will discuss some archetypal examples to elaborate our construction.

\section{GDWs in the \texorpdfstring{$\mathbb{Z}_2$}{} Toric code phase \label{sec: GDWs in Toric code phase}}

In this section, we produce all the GDWs that separate the two toric code phases. Notably, the  $e$-$m$ exchanging GDWs can be realized in our GDW construction framework, without inserting additional $\sigma$ defects or introducing lattice dislocalization. This method is also applicable to studying $e$-$m$ exchanging GDWs between the different LW models, such as those with input UFCs $\mathcal{R}ep(\mathbb{Z}_n)$ and $\mathcal{V}ec(\mathbb{Z}_n)$.

\subsection{ \texorpdfstring{$\mathcal{C}_1 = \mathcal{C}_2 = \mathcal{R}ep(\mathbb{Z}_2)$}{}}
\label{subsubsection: Hamiltonians of LW Z2}
Here we consider two $\mathbb{Z}_2$ LW models separated by a GDW. 
The input fusion category here is $\mathcal{R}ep(\mathbb{Z}_2)$, which has two self-dual simple objects $1$ and $\psi$, with $d_1=d_\psi=1$. The fusion rules are $\delta_{111}= \delta_{1\psi\psi}= 1$. The 6$j$-symbols are $G^{ijm}_{kln}=\delta_{ijm}\delta_{klm^*}\delta_{jkn^*}\delta_{inl}$. Table \ref{tab:solutions of domain walls in Toric code} records all possible triples $(A_1, A_2, \eta)$:

\begin{table}[htbp]
    \centering
    \renewcommand\arraystretch{1.5}
    \begin{tabular}{|c|l|l|}
        \hline
        \diagbox{$A_1$}{$A_2$} & \quad \quad \quad \quad 1 & \quad \quad \quad \quad \quad \ $1\oplus\psi$ \\
        \hline 
         1& {\footnotesize (1)} $\eta_{11}=1$ & {\footnotesize (2)} $\eta_{11}=\eta_{1\psi}= \frac{1}{\sqrt{2}}$ \\
         \hline 
         \multirow{2}{*}{$1\oplus \psi$} & \multirow{2}{*}{{\footnotesize (3)} $\eta_{11}=\eta_{\psi 1}=\frac{1}{\sqrt{2}}$} & {\footnotesize (4)} $\eta_{11}=\eta_{1\psi}=\eta_{\psi 1}=\eta_{\psi \psi}=\frac{1}{2}$ \\ \cline{3-3}
          & & {\footnotesize (5)} $\eta_{11}=\eta_{\psi \psi}=\frac{1}{\sqrt{2}}$ \\
         \hline
    \end{tabular}
    \caption{Solutions $\eta$ for different algebra objects $A_1$ and $A_2$ in the $\mathbb{Z}_2$ fusion category. Algebras $A_1$ and $A_2$ both happen to be Frobenius algebras, respectively with multiplications $f_{ijk}=\delta_{ijk}$ and $g_{mnl}=\delta_{mnl}$ where $i,j,k\in A_1$ and $m,n,l\in A_2$.}
    \label{tab:solutions of domain walls in Toric code}
\end{table}
By substituting the five solutions listed in Table \ref{tab:solutions of domain walls in Toric code} into the defining equation \eqref{eq: DW plaquette op} of DW plaquette operators
\begin{equation}
    B_p^\text{DW}=\sum_{s\in A_1, t\in A_2} v_s v_t (\eta_{st})^2 B^{st}_p,
\end{equation}
we derive five distinct DW plaquette operators: 
\begin{equation}
    \begin{aligned}
    B_p^{\text{DW}_1} & =B_p^{11}, \quad  B_p^{\text{DW}_2}=\frac{B_p^{11}+B_p^{1\psi}}{2}, \quad B_p^{\text{DW}_3}=\frac{B_p^{11}+B_p^{\psi 1}}{2}, \\ B_p^{\text{DW}_4}&=\frac{B_p^{11}+B_p^{1\psi}+B_p^{\psi 1}+B_p^{\psi \psi}} {4}, \quad B_p^{\text{DW}_5}=\frac{B_p^{11}+B_p^{\psi \psi}}{2}.
\end{aligned}
\label{eq: five Hamiltonians for LW Z2}
\end{equation}

We observe that the fifth DW plaquette operator $B_p^{\text{DW}_5}$ coincides with the bulk plaquette operator $B_p^\text{bulk}$, so the corresponding GDW is trivial: The system is just a single $\mathbb{Z}_2$ toric code phase without any domain wall. 
The other four DW plaquette operators correspond to nontrivial GDWs. To understand these GDWs, we can decompose these DW plaquette operators as
\begin{equation}
\begin{aligned}
        B_p^\text{DW}&= \sum_{i\in A_1, j\in A_2} \eta_{ij}^2 v_i v_j B_p^{ij} \\
        &=\sum_{i\in A_1, j\in A_2} \frac{1}{d_{A_1}} \frac{1}{d_{A_2}} \delta_{i\in A_1} \delta_{j\in A_2} v_i v_j B_p^{ij}\\
        &=\sum_{i\in A_1, j\in A_2} \frac{1}{d_{A_1}} \frac{1}{d_{A_2}}  (v_i B_{p_l}^{i}  \otimes v_j B_{p_r}^{j}) \\
        &= (\frac{1}{d_{A_1}}\sum_{i\in A_1} v_i B_{p_l}^i) \otimes (\frac{1}{d_{A_2}}\sum_{j\in A_2} v_j B_{p_r}^j)
        &=: B^\text{DW}_{p_l}  \otimes B^\text{DW}_{p_r},
        \label{eq: half-plaquette operators}
\end{aligned}
\end{equation}
In the third equality above, we rewrite $B_p^{st}$ as $B_{p_l}^s \otimes B_{p_l}^t$, where $p_l$ ($p_r$) labels the left (right) half of the plaquette $p$ of the joining points. This decomposition is possible because the joining function can be expressed as $\eta_{ij}= \sqrt{d_{A_1}^{-1}}\sqrt{d_{A_2}^{-1}} \delta_{i\in A_1} \delta_{j\in A_2}$. 

According to Ref. \cite{hu2018boundary}, a GB of the tailed LW model is specified by a Frobenius algebra object $A$ of the model's input UFC. Such a GB consists of half plaquettes with plaquette operators defined as
\begin{equation}
    B_p^\text{GB}=\frac{1}{d_A} \sum_{i\in A} v_i B_p^{i},
\end{equation}
which agrees with the form of the two half-plaquette operators $B^\text{DW}_{p_l}$ and $B^\text{DW}_{p_r}$ in the decomposition \eqref {eq: half-plaquette operators}. This is natural because both $A_1, A_2$ in Eq. \eqref{eq: half-plaquette operators} are Frobenius algebras. As such, a DW plaquette operator $B_p^\text{DW}$ is decomposed into the tensor product of two half-plaquette operators, as in Eq. \eqref {eq: half-plaquette operators}.

In view of this decomposition, the GDW with the first DW plaquette operator $B_p^{\text{DW}_1}$ is understood as joining the two GBs of the toric code phases on both sides. Each GB is an $m$-boundary\footnote{Note that the convention of the $e$($m$)-boundary may differ from those in some literatures, where the $e$($m$)-boundary means $e$($m$) is confined at the GB.}, where anyon $m$ condenses. Likewise, the GDW with the fourth DW plaquette operator $B_p^{\text{DW}_4}$ glues the two GBs of the toric code phases on both sides, each of which is an $e$-boundary, where anyon $e$ condenses. 

The GDW with the second DW plaquette operator $B_p^{\text{DW}_2}$ joins an $m$-boundary of the $\mathbb{Z}_2$ LW model on the left and the $e$-boundary on the right. This GDW facilitates the transformation of a condensed anyon $m$ on the left to a condensed anyon $e$ on the right. In contrast, the GDW with the third DW plaquette operator $B_p^{\text{DW}_2}$ joins an $e$-boundary of the $\mathbb{Z}_2$ LW model on the left and the $m$-boundary on the right. These two GDWs can be distinguished by the following: When we weld the two composite systems vertically on top of each other, a junction would have to appear where the two GDWs are welded. The junction arises because we are welding an $m$-boundary with an $e$-boundary and vice versa\cite{wang2022extend}.

Our model yields 5 GDWs within the $\mathbb{Z}_2$ LW model. Nevertheless, it is known that the $\mathbb{Z}_2$ LW model possesses another GDW, crossing which an anyon $e$ ($m$) on one side becomes an anyon $m$ ($e$) on the other side\cite{Kitaev-Kong-2012,Hung-Wan-ADE-2015,bombin2010topological}. In our framework, this $e-m$ exchanging GDW separates the $\mathbb{Z}_2$ LW models with different input fusion categories of $\mathbb{Z}_2$ symmetry: one being $\mathcal{R}ep(\mathbb{Z}_2)$ and the other $\mathcal{V}ec(\mathbb{Z}_2)$.

\subsection{ \texorpdfstring{$\mathcal{C}_1 = \mathcal{R}ep(\mathbb{Z}_2)$}{} and \texorpdfstring{$\mathcal{C}_2 = \mathcal{V}ec(\mathbb{Z}_2)$}{}: Realizing e-m exchange
}\label{sec: e-m exchanging GDW in TC}

Let us consider the GDW that separates a
$\mathbb{Z}_2$ LW model with input $\mathcal{R}ep(\mathbb{Z}_2)$ and another $\mathbb{Z}_2$ model with input $\mathcal{V}ec(\mathbb{Z}_2)$. Note that $\mathcal{C}_2$ is not a subcategory of $\mathcal{C}_1$. 
The two simple objects of $\mathcal{V}ec(\mathbb{Z}_2)$ are equivalent to the two irreducible $A$-bimodules $N_0,N_1$ of $\mathcal{R}ep(\mathbb{Z}_2)$, where $A$ is the nontrivial Frobenius algebra object $1\oplus\psi$ in $\mathcal{R}ep(\mathbb{Z}_2)$. 
More precisely, $N_0=(1\oplus\psi, P_{N_0})$ with trivial bimodule action tensors and $N_1=(1\oplus\psi, P_{N_1})$ with nontrivial bimodule action tensors given by \eqref{eq: AAbimodulesM0-A=1+psi} and \eqref{eq: AAbimodulesM1-A=1+psi} in Appendix.\ref{Appendix: A1-bimodules-Z2}.

The fusion rules of $N_0$ and $N_1$ are $\delta_{N_0 N_0 N_0}=\delta_{N_0 N_1 N_1}=1$. The quantum dimensions for bimodules are $d_{N_0}=d_{N_1}=1$. Moreover, 6j-symbols are given by $G_{N_k N_l N_n}^{N_i N_j N_m}= \delta_{N_i N_j N_m} \delta_{N_k N_l N_m} \delta_{N_j N_k N_n} \delta_{N_i N_n N_l}$. 
Table \ref{tab:solutions of domain walls in Toric code2} records all possible triples $(A_1, A_2, \eta)$:
\begin{table}[htbp]
    \centering
    \renewcommand\arraystretch{1.5}
    \begin{tabular}{|c|l|l|}
        \hline
        \diagbox{$A_1$}{$A_2$} & \quad \quad \quad \quad $N_0$ & \quad \quad \quad \quad \quad \quad $N_0\oplus N_1$ \\
        \hline 
         1& {\footnotesize (2)} $\eta_{1N_0}=1$ & {\footnotesize (1)} $\eta_{1N_0}=\eta_{1 N_1}= \frac{1}{\sqrt{2}}$ \\
         \hline 
         \multirow{2}{*}{$1\oplus \psi$} & \multirow{2}{*}{{\footnotesize (4)} $\eta_{1N_0}=\eta_{\psi N_0}=\frac{1}{\sqrt{2}}$} & {\footnotesize (3)} $\eta_{1N_0}=\eta_{1N_1}=\eta_{\psi N_0}=\eta_{\psi N_1}=\frac{1}{2}$ \\ \cline{3-3}
          & & {\footnotesize (6)} $\eta_{1N_0}=\eta_{\psi N_1}=\frac{1}{\sqrt{2}}$ \\
         \hline
    \end{tabular}
    \caption{Solutions $\eta$ for different algebra objects $A_1$ in $\mathcal{R}ep(\mathbb{Z}_2)$ and $A_2$ in $\mathcal{V}ec(\mathbb{Z}_2)$. Algebras $A_1$ in $\mathcal{R}ep(\mathbb{Z}_2)$ and $A_2$ in $\mathcal{V}ec(\mathbb{Z}_2)$ have multiplications $f_{ijk}=\delta_{ijk}$ and $g_{N_i N_j N_k}=\delta_{N_i N_j N_k}$, respectively, where $i,j,k\in A_1$ and $N_i,N_j,N_k\in A_2$.}
    \label{tab:solutions of domain walls in Toric code2}
\end{table}

We label the solutions in Table \ref{tab:solutions of domain walls in Toric code2} according to the types of GDWs, as the first four solutions (1)-(4) in Table \ref{tab:solutions of domain walls in Toric code2} have one-to-one correspondence with the first four solutions in Table \ref{tab:solutions of domain walls in Toric code}, as shown below:
\begin{equation}
    \begin{aligned}
        (1)\ & \eta_{11}=1 & \longleftrightarrow \quad  \ & \eta_{1N_0}=\eta_{1 N_1}= 1/{\sqrt{2}} \\  
        (2)\ & \eta_{11}=\eta_{1\psi}= 1/{\sqrt{2}} & \longleftrightarrow \quad \ &\eta_{1N_0}=1 \\  
        (3)\ & \eta_{11}=\eta_{\psi 1}=1/{\sqrt{2}} & \longleftrightarrow \quad \ & \eta_{1N_0}=\eta_{1N_1}=\eta_{\psi N_0}=\eta_{\psi N_1}=1/{2} \\
        (4)\ & \eta_{11}=\eta_{1\psi}=\eta_{\psi 1}=\eta_{\psi \psi}=1/{2} & \longleftrightarrow \quad \  & \eta_{1N_0}=\eta_{\psi N_0}=1/{\sqrt{2}}
    \end{aligned}
\end{equation}

Here comes the question: What is the GDW corresponding to solution (6) in Table \ref{tab:solutions of domain walls in Toric code2}? Let us construct an interdomain ribbon operator that transports an anyon in the $\mathcal{R}ep(\mathbb{Z}_2)$-model to another anyon in the $\mathcal{V}ec(\mathbb{Z}_2)$-model. The interdomain ribbon operator must commute with the DW plaquette operators $B_p^\text{DW}$ to ensure that it is does not create any excitations in the GDW. Moreover, an observation from solution (6) is that the gluing function $\eta$ always connects the DOF $1$ ($\psi$) in the $\mathcal{R}ep(\mathbb{Z}_2)$-model and the DOF $N_0$ ($N_1$) in the $\mathcal{V}ec(\mathbb{Z}_2)$-model. Therefore, upon crossing the GDW, a $\mathcal{R}ep(\mathbb{Z}_2)$-model anyon whose tail DOF is $1$ ($\psi$) would become a $\mathcal{V}ec(\mathbb{Z}_2)$-model anyon, whose tail DOF is $N_0$ ($N_1$).

According to the ribbon operators in the tailed LW model~\cite{hu2018full}, the DOF of the tail for anyon $e$ in the $\mathcal{R}ep(\mathbb{Z}_2)$-model is $\psi$, while that for $m$ in the $\mathcal{V}ec(\mathbb{Z}_2)$-model is $N_1$. 
We can construct the following operator, which can create $e$ in the $\mathcal{R}ep(\mathbb{Z}_2)$-model and $m$ in the $\mathcal{V}ec(\mathbb{Z}_2)$-model:

\begin{equation}
    W^{e-m}_{E_1,E_2} \Bigg| \tikzset{every picture/.style={line width=0.75pt}} 
\begin{tikzpicture}[x=0.75pt,y=0.75pt,yscale=-1,xscale=1, baseline=(XXXX.south) ]
\path (0,121);\path (152.69009399414062,0);\draw    ($(current bounding box.center)+(0,0.3em)$) node [anchor=south] (XXXX) {};
\draw  [draw opacity=0][fill={rgb, 255:red, 208; green, 2; blue, 27 }  ,fill opacity=0.34 ][dash pattern={on 4.5pt off 4.5pt}][line width=0.75]  (5.76,6.73) -- (41.65,6.73) -- (41.65,116.59) -- (5.76,116.59) -- cycle ;
\draw  [draw opacity=0][fill={rgb, 255:red, 74; green, 144; blue, 226 }  ,fill opacity=0.42 ][dash pattern={on 4.5pt off 4.5pt}][line width=0.75]  (110.77,6.73) -- (145.76,6.73) -- (145.76,116.59) -- (110.77,116.59) -- cycle ;
\draw  [draw opacity=0][fill={rgb, 255:red, 128; green, 128; blue, 128 }  ,fill opacity=0.1 ][dash pattern={on 4.5pt off 4.5pt}][line width=0.75]  (41.65,6.8) -- (110.77,6.8) -- (110.77,116.04) -- (41.65,116.04) -- cycle ;
\draw    (110.77,6.73) -- (110.77,116.04) ;
\draw    (41.65,6.73) -- (41.65,116.04) ;
\draw [color={rgb, 255:red, 208; green, 2; blue, 27 }  ,draw opacity=1 ]   (42.35,97.65) -- (72.16,97.65) ;
\draw [color={rgb, 255:red, 33; green, 55; blue, 191 }  ,draw opacity=1 ]   (76.37,97.65) -- (110.19,97.65) ;
\draw [color={rgb, 255:red, 208; green, 2; blue, 27 }  ,draw opacity=1 ]   (42.35,25) -- (72.15,25) ;
\draw [color={rgb, 255:red, 33; green, 55; blue, 191 }  ,draw opacity=1 ]   (80.57,25) -- (110.19,25) ;
\draw    (110.77,59.46) -- (145.25,59.46) ;
\draw    (5.62,61.46) -- (41.65,61.46) ;
\draw  [fill={rgb, 255:red, 255; green, 255; blue, 255 }  ,fill opacity=1 ] (72.15,20.79) -- (80.57,20.79) -- (80.57,29) -- (72.15,29) -- cycle ;
\draw  [fill={rgb, 255:red, 255; green, 255; blue, 255 }  ,fill opacity=1 ] (72.16,93.45) -- (80.58,93.45) -- (80.58,101.65) -- (72.16,101.65) -- cycle ;
\draw (48.21,10.66) node [anchor=north west][inner sep=0.75pt]  [font=\small]  {$a_{1}$};
\draw (89.99,8.93) node [anchor=north west][inner sep=0.75pt]  [font=\small]  {$b_{1}$};
\draw (48.89,100.48) node [anchor=north west][inner sep=0.75pt]  [font=\small]  {$a_{2}$};
\draw (92.62,98.26) node [anchor=north west][inner sep=0.75pt]  [font=\small]  {$b_{2}$};
\draw (42.81,34.7) node [anchor=north west][inner sep=0.75pt]  [font=\small]  {$i_{1}$};
\draw (42.65,69.16) node [anchor=north west][inner sep=0.75pt]  [font=\small]  {$i_{2}$};
\draw (18.6,64.32) node [anchor=north west][inner sep=0.75pt]  [font=\small]  {$i_{3}$};
\draw (97.21,34.05) node [anchor=north west][inner sep=0.75pt]  [font=\small]  {$j_{1}$};
\draw (96.77,69.16) node [anchor=north west][inner sep=0.75pt]  [font=\small]  {$j_{2}$};
\draw (126.51,62.27) node [anchor=north west][inner sep=0.75pt]  [font=\small]  {$j_{3}$};
\draw (68.35,31.42) node [anchor=north west][inner sep=0.75pt]  [font=\footnotesize]  {$M_{0}$};
\draw (22.68,33.88) node [anchor=north west][inner sep=0.75pt]  [font=\small]  {$E_{1}$};
\draw (112.19,34.7) node [anchor=north west][inner sep=0.75pt]  [font=\small]  {$E_{2}$};
\draw (70,54.88) node [anchor=north west][inner sep=0.75pt]  [font=\small]  {$p$};
\end{tikzpicture}
 \Vast\rangle = \Bigg| \tikzset{every picture/.style={line width=0.75pt}} 
\begin{tikzpicture}[x=0.75pt,y=0.75pt,yscale=-1,xscale=1, baseline=(XXXX.south) ]
\path (0,124);\path (154.8528594970703,0);\draw    ($(current bounding box.center)+(0,0.3em)$) node [anchor=south] (XXXX) {};
\draw  [draw opacity=0][fill={rgb, 255:red, 208; green, 2; blue, 27 }  ,fill opacity=0.34 ][dash pattern={on 4.5pt off 4.5pt}][line width=0.75]  (5.65,6.01) -- (42.31,6.01) -- (42.31,118.56) -- (5.65,118.56) -- cycle ;
\draw  [draw opacity=0][fill={rgb, 255:red, 74; green, 144; blue, 226 }  ,fill opacity=0.42 ][dash pattern={on 4.5pt off 4.5pt}][line width=0.75]  (113.43,5.38) -- (149.65,5.38) -- (149.65,117.92) -- (113.43,117.92) -- cycle ;
\draw  [draw opacity=0][fill={rgb, 255:red, 128; green, 128; blue, 128 }  ,fill opacity=0.1 ][dash pattern={on 4.5pt off 4.5pt}][line width=0.75]  (42.31,6.01) -- (113.43,6.01) -- (113.43,117.92) -- (42.31,117.92) -- cycle ;
\draw    (113.43,5.94) -- (113.43,117.92) ;
\draw    (42.31,5.94) -- (42.31,117.92) ;
\draw [color={rgb, 255:red, 208; green, 2; blue, 27 }  ,draw opacity=1 ]   (43,98.64) -- (74.19,98.64) ;
\draw [color={rgb, 255:red, 33; green, 55; blue, 191 }  ,draw opacity=1 ]   (82.62,98.64) -- (112.84,98.64) ;
\draw [color={rgb, 255:red, 208; green, 2; blue, 27 }  ,draw opacity=1 ]   (43,27.26) -- (73.19,27.26) ;
\draw [color={rgb, 255:red, 33; green, 55; blue, 191 }  ,draw opacity=1 ]   (81.62,27.26) -- (112.84,27.26) ;
\draw    (113.43,78.77) -- (149.15,78.77) ;
\draw    (6.03,78.77) -- (42.06,78.77) ;
\draw    (77.41,52.55) -- (113.24,52.55) ;
\draw    (51.63,52.76) -- (73.19,52.76) ;
\draw    (121.43,52.62) -- (132.42,52.62) ;
\draw    (23.3,52.14) -- (34.08,52.14) ;
\draw  [fill={rgb, 255:red, 255; green, 255; blue, 255 }  ,fill opacity=1 ] (34.47,46.76) .. controls (34.47,44.9) and (35.98,43.39) .. (37.84,43.39) -- (47.93,43.39) .. controls (49.79,43.39) and (51.3,44.9) .. (51.3,46.76) -- (51.3,57.7) .. controls (51.3,59.56) and (49.79,61.07) .. (47.93,61.07) -- (37.84,61.07) .. controls (35.98,61.07) and (34.47,59.56) .. (34.47,57.7) -- cycle ;
\draw  [fill={rgb, 255:red, 255; green, 255; blue, 255 }  ,fill opacity=1 ] (104.99,47.36) .. controls (104.99,45.5) and (106.49,43.99) .. (108.35,43.99) -- (118.45,43.99) .. controls (120.31,43.99) and (121.82,45.5) .. (121.82,47.36) -- (121.82,58.3) .. controls (121.82,60.16) and (120.31,61.67) .. (118.45,61.67) -- (108.35,61.67) .. controls (106.49,61.67) and (104.99,60.16) .. (104.99,58.3) -- cycle ;
\draw  [fill={rgb, 255:red, 255; green, 255; blue, 255 }  ,fill opacity=1 ] (73.19,23.02) -- (81.62,23.02) -- (81.62,31.23) -- (73.19,31.23) -- cycle ;
\draw  [fill={rgb, 255:red, 255; green, 255; blue, 255 }  ,fill opacity=1 ] (73.19,49.02) -- (81.62,49.02) -- (81.62,57.23) -- (73.19,57.23) -- cycle ;
\draw  [fill={rgb, 255:red, 255; green, 255; blue, 255 }  ,fill opacity=1 ] (74.19,95.02) -- (82.62,95.02) -- (82.62,103.23) -- (74.19,103.23) -- cycle ;
\draw (48.86,12.71) node [anchor=north west][inner sep=0.75pt]  [font=\small]  {$a_{1}$};
\draw (92.64,9.97) node [anchor=north west][inner sep=0.75pt]  [font=\small]  {$b_{1}$};
\draw (49.55,99.3) node [anchor=north west][inner sep=0.75pt]  [font=\small]  {$a_{2}$};
\draw (95.27,96.69) node [anchor=north west][inner sep=0.75pt]  [font=\small]  {$b_{2}$};
\draw (54.94,40.23) node [anchor=north west][inner sep=0.75pt]  [font=\scriptsize]  {$\psi $};
\draw (85.7,40.34) node [anchor=north west][inner sep=0.75pt]  [font=\scriptsize]  {$N_{1}$};
\draw (9.36,43.95) node [anchor=north west][inner sep=0.75pt]  [font=\scriptsize]  {$\psi $};
\draw (130.79,45.27) node [anchor=north west][inner sep=0.75pt]  [font=\scriptsize]  {$N_{1}$};
\draw (35.48,45.44) node [anchor=north west][inner sep=0.75pt]  [font=\footnotesize]  {$z^{e}$};
\draw (104.4,45.84) node [anchor=north west][inner sep=0.75pt]  [font=\footnotesize]  {$z^{m}$};
\draw (4.12,63.61) node [anchor=north west][inner sep=0.75pt]  [font=\small]  {$i_{3}$};
\draw (138.03,61.77) node [anchor=north west][inner sep=0.75pt]  [font=\small]  {$j_{3}$};
\draw (69.95,60.95) node [anchor=north west][inner sep=0.75pt]  [font=\footnotesize]  {$M_{0}$};
\draw (27.65,83.27) node [anchor=north west][inner sep=0.75pt]  [font=\small]  {$i_{2}$};
\draw (113.77,82.27) node [anchor=north west][inner sep=0.75pt]  [font=\small]  {$j_{2}$};
\draw (28,27.96) node [anchor=north west][inner sep=0.75pt]  [font=\small]  {$i_{1}$};
\draw (28.81,62.7) node [anchor=north west][inner sep=0.75pt]  [font=\small]  {$i_{1}$};
\draw (114.21,27.05) node [anchor=north west][inner sep=0.75pt]  [font=\small]  {$j_{1}$};
\draw (113.21,62.05) node [anchor=north west][inner sep=0.75pt]  [font=\small]  {$j_{1}$};
\end{tikzpicture}
\Vast\rangle.
    \label{eq: e-m ribbon op}
\end{equation}
Here, $E_1, E_2$ are the edges that intersects with the ribbon operator, whose DOFs are labeled by $i_1,i_2$, respectively. The boxes labeled by $z^e$ and $z^m$ are half-braiding tensors correspond to $e$ in the $\mathcal{R}ep(\mathbb{Z}_2)$-model and $m$ in the $\mathcal{V}ec(\mathbb{Z}_2)$-model, and their data are given in Appendix \ref{Appendix: Half-braiding tensors-Z2} and \ref{Appendix: Half-braiding tensors2-Z2}. This ribbon operator commutes with the DW plaquette operator $B_p^{\text{DW}_6}=\frac{B_{p}^{1 N_0} + B_{p}^{\psi N_1}}{2}$, which has been proven in Appendix \ref{proof: Bp commutes with W}. Thus, this operator causes no excitation in the DW plaquette, and it is exactly the desired interdomain ribbon operator.
Likewise, we can construct the DW-crossing ribbon operators $W^{m-e}_{E_1,E_2}$, $W^{1-1}_{E_1,E_2}$ and $W^{\epsilon-\epsilon}_{E_1,E_2}$, all of which commute with $B_p^{\text{DW}_6}$. According to the proof in Appendix \ref{proof: Bp commutes with W}, there are no other ribbon operators in the form of $W^{J_1-J_2}_{E_1,E_2}$. Therefore, anyons $1,\epsilon,e,m$ in the $\mathcal{R}ep(\mathbb{Z}_2)$-model would become $1,\epsilon,m,e$ in the $\mathcal{V}ec(\mathbb{Z}_2)$-model, respectively. So the GDW corresponding to Solution (6) is the $e$-$m$ exchanging GDW.

Based on the analysis above, we can see that the GDW corresponding to Solution (6) realizes $e$-$m$ exchange. Moreover, the results of $\AAeta$-bimodules corresponding to Solution (6) and other four solutions in Table \ref{tab:solutions of domain walls in Toric code2} are provided in Appendix. 
\ref{Appendix: A1-bimodules-Z2}.

In summary, the $\Z_2$ toric code phase possesses six physically distinguishable GDWs:
 \begin{enumerate}[label=\Roman*.]
     \item $(A_1=1,A_2=1,\eta_{11}=1)$ characterizes a GDW that is equivalent to joining two $m$-condensed boundaries of the toric code.
    
    \item $(A_1=1,A_2=1\oplus\psi,\eta_{11}=\eta_{1\psi}=\frac{1}{2})$ characterizes a GDW that is equivalent to joining a left $m$-condensed boundary and a right $e$-condensed boundary of the toric code.
    
    \item $(A_1=1\oplus\psi, A_2=1, \eta_{11}=\eta_{\psi 1}=\frac{1}{2})$ characterizes a GDW that is equivalent to joining a left $e$-condensed boundary and a right $m$-condensed boundary of the toric code.
    
    \item $(A_1=1\oplus\psi, A_2=1\oplus\psi, \eta_{11}=\eta_{1\psi}=\eta_{\psi 1}=\eta_{\psi\psi}=\frac{1}{2})$ characterizes a GDW that is equivalent to joining two $e$-condensed boundaries of the toric code.
    
    \item $(A_1=1\oplus\psi, A_2=1\oplus\psi, \eta_{11}=\eta_{\psi\psi}=\frac{1}{\sqrt{2}})$ characterizes the trivial GDW in the toric code.
    \item $(A_1=1\oplus\psi, A_2=N_0 \oplus N_1, \eta_{1 N_0}=\eta_{\psi N_1}=\frac{1}{2})$ characterizes a GDW that realizes $e$-$m$ exchange in the toric code.
\end{enumerate}

\section{GDWs in the doubled Ising phase}  \label{sec: GDWs in the doubled Ising}
In this section, we consider GDWs in the doubled Ising phase. We find that our construction can derive all GDWs at the microscopic (input) level. However, from the classification at the macroscopic (output) level, some GDWs may lead to equivalent results.

The input fusion category of the doubled Ising LW model is the Ising fusion category, which has objects $\{1,\sigma ,\psi \}$ with fusion rules $\psi\times\psi=1, \sigma\times\sigma=1+\psi, \sigma\times\psi=\sigma$, so the quantum dimensions for each simple objects are $d_1=d_{\psi}=1,d_{\sigma}=\sqrt{2}$, and the fusion coefficients are
\begin{equation}
    \delta_{111}= \delta_{1\sigma\sigma}= \delta_{1\psi\psi}= \delta_{\sigma\sigma\psi}= 1.
\end{equation}
The nonvanishing 6$j$-symbols are
\begin{equation}
\begin{aligned}
    G_{111}^{111}& =1, G_{\psi\psi\psi}^{111}=1, G_{1\psi\psi}^{1\psi\psi}=1, G_{\sigma\sigma\sigma}^{111}=\frac{1}{\sqrt[4]{2}}, \\
    G_{1\sigma\sigma}^{1\sigma\sigma}& =\frac{1}{\sqrt{2}}, G_{\sigma\psi\psi}^{1\sigma\sigma}=\frac{1}{\sqrt[4]{2}}, G_{\psi\sigma\sigma}^{1\sigma\sigma}=\frac{1}{\sqrt{2}}, G_{\sigma\sigma\psi}^{\sigma\sigma\psi}=-\frac{1}{\sqrt{2}}.
\end{aligned}
\end{equation}
Solutions of $\eta$ to Eq. \eqref{eq: domain wall Bp commutativity} are listed below:

\begin{table}[htbp]
    \centering
    \renewcommand\arraystretch{1.5}
    \begin{tabular}{|c|l|l|l|}
        \hline
        \diagbox{$A_1$}{$A_2$} & \quad \quad \quad \quad 1 & \quad \quad \quad \quad \quad $1\oplus \psi$ & \quad \quad \quad $1\oplus \psi \oplus \sigma$ \\
        \hline 
         1 & {\footnotesize (1)} $\eta_{11}=1$ & {\footnotesize (2)} $\eta_{11}=\eta_{1\psi}=\frac{1}{\sqrt{2}}$ & \quad \quad \quad \quad \quad - \\
         \hline 
         \multirow{2}{*}{$1\oplus \psi$} & \multirow{2}{*}{{\footnotesize (3)} $\eta_{11}=\eta_{\psi 1}=\frac{1}{\sqrt{2}}$} & {\footnotesize (4)} $\eta_{11}=\eta_{1\psi}=\eta_{\psi 1}=\eta_{\psi\psi}=\frac{1}{2}$ & \quad \quad \quad \quad \quad \multirow{2}{*}{-} \\
         \cline{3-3}
          &  & {\footnotesize (5)} $\eta_{11}=\eta_{\psi\psi}=\frac{1}{\sqrt{2}}$ & \\ 
         \hline
         \multirow{1}{*}{$1\oplus \psi \oplus \sigma$} & \quad \quad \quad \quad \multirow{1}{*}{-} & \quad \quad \quad \quad \quad \quad \multirow{1}{*}{-} & {\footnotesize (6)} $\eta_{11}=\eta_{\psi\psi}=\eta_{\sigma\sigma}=\frac{1}{2}$ \\
         \hline
    \end{tabular}
    \caption{Solutions to $\eta$ for different $A_1$ and $A_2$ in the LW Ising model. In this case, $f_{ijk}=\delta_{ijk},g_{mnl}=\delta_{mnl}$ where $i,j,k\in A_1$ and $m,n,l\in A_2$.}
    \label{tab:solutions of domain walls in the Ising LW}
\end{table}
To understand which GDWs are characterized by the solutions in Table \ref{tab:solutions of domain walls in the Ising LW},
we first note that the gluing functions in Solutions (1)-(4) satisfy the form: $\eta_{ab}=\sqrt{d_{A_1}^{-1}} \sqrt{d_{A_2}^{-1}} \delta_{a\in A_1} \delta_{b \in A_2}$. As a result, Solutions (1)-(4) characterize GDWs that result from gluing the two GBs of the doubled Ising. Moreover, $A=1$ and $A=1\oplus\psi$ are Frobenius algebras that characterize the two Morita-equivalent GBs of the doubled Ising  \cite{hu2018boundary}, so the four GDWs characterized by Solutions (1)-(4) are physically equivalent. 
The corresponding $\AAeta$-bimodules are given in Appendix. \ref{Appendix: A1-bimodules-Ising}.

Solution (5) characterizes a $\psi\bar{\psi}$-condensation induced GDW. To illustrate this, we can construct an interdomain ribbon operator as follows:
\begin{equation}
    W^{\psi\bar{\psi}-1\bar{1}}_{E_1,E_2} \Bigg| \input{images/DW_excitations/fig__interdomain_ribbon_psipsi-11_1} \Vast\rangle = \Bigg| \input{images/DW_excitations/fig__interdomain_ribbon_psipsi-11_2}\Vast\rangle.
    \label{eq: psipsi-11 ribbon op}
\end{equation}

This operator describes the process of $\psi\bar{\psi}$-condensation across the GDW. As we will see in the next section, this condensation occurs between the doubled Ising phase and the toric code phase, thus Solution (5) describes a $(1+1)$-dimensional defect formed by shrinking the toric code phase.

In conclusion, the six solutions in Table \ref{tab:solutions of domain walls in the Ising LW} lead to only three physically inequivalent GDWs in the doubled Ising topological phase:
\begin{enumerate}[label=\Roman*.]
    \item The first four solutions characterize a common GDW that is derived by gluing two Morita-equivalent GBs of the doubled Ising.
    
    \item $(A_1=1,A_2=1\oplus\psi,\eta_{11}=\eta_{\psi\psi}=\frac{1}{\sqrt{2}})$ characterizes the $\psi\bar{\psi}$-condensing GDW in the doubled Ising. 

    \item $(A_1=1,A_2=1\oplus\psi\oplus\sigma,\eta_{11}=\eta_{\psi\psi}=\eta_{\sigma\sigma}=\frac{1}{2})$ characterizes the trivial GDW in the doubled Ising. 
\end{enumerate}

\section{GDWs between the doubled Ising and Toric code phases} \label{sec: GDWs in DI-TC}
Here, we compute GDWs separating the doubled Ising and the toric code. There are five solutions of the joining function $\eta$ given different $A_1$ and $A_2$, as listed in table~\ref{tab:solutions of domain walls in Doubled Ising-Toric code}. 
\begin{table}[htbp]
    \centering
    \renewcommand\arraystretch{1.5}
    \begin{tabular}{|c|l|l|}
        \hline
        \diagbox{$A_1$}{$A_2$} & \quad \quad \quad \quad 1 & \quad \quad \quad \quad \quad \ $1\oplus\psi$ \\
        \hline 
         1& {\footnotesize (1)} $\eta_{11}=1$ & {\footnotesize (2)} $\eta_{11}=\eta_{1\psi}= \frac{1}{\sqrt{2}}$ \\
         \hline 
         \multirow{2}{*}{$1\oplus \psi$} & \multirow{2}{*}{{\footnotesize (3)} $\eta_{11}=\eta_{\psi 1}=\frac{1}{\sqrt{2}}$} & {\footnotesize (4)} $ \eta_{11}=\eta_{1\psi}=\eta_{\psi 1}=\eta_{\psi \psi}=\frac{1}{2}$ \\ \cline{3-3}
          & & {\footnotesize (5)} $\eta_{11}=\eta_{\psi \psi}=\frac{1}{\sqrt{2}}$ \\
         \hline
    \end{tabular}
        \caption{Solutions to $\eta$ for different $A_1$ and $A_2$ for the GDWs between Ising LW model and $\mathbb{Z}_2$ LW models. In this case, $f_{ijk}=\delta_{ijk},g_{mnl}=\delta_{mnl}$ where $i,j,k\in A_1$ and $m,n,l\in A_2$, with $A_1$ and $A_2$ the Frobenius algebras for the Ising LW and the $\mathbb{Z}_2$ LW model, respectively.}
    \label{tab:solutions of domain walls in Doubled Ising-Toric code}
\end{table}

The first four solutions labeled by (1)-(4) characterize GDWs that are obtained by gluing the GBs of the doubled Ising and toric code. Since the two GBs of the doubled Ising are equivalent~\cite{hu2018boundary}, solutions (1) and (3) characterize the physically equivalent GDW that joins the GB of the doubled Ising and the $m$-boundary of the toric code. Similarly, Solutions (2) and (4) characterize the GDW that joins the GB of doubled Ising and the $e$-boundary of the toric code. 

Solution (5) in Table \ref{tab:solutions of domain walls in Doubled Ising-Toric code}  corresponds to a phase transition process. In this process, the right part of the Levin-Wen (LW) model, initially in the doubled Ising phase, undergoes $\psi\bar{\psi}$-condensation and transitions into the toric code phase~\cite{zhao2023characteristic}. As a result, a GDW emerges at the interface between the toric code phase (on the right) and the remaining doubled Ising phase (on the left), which is precisely characterized by Solution (5).

In summary, the doubled Ising and $\Z_2$ topological phases can be separated by three physically distinct GDWs:
\begin{enumerate}[label=\Roman*.]
 
    \item $(A_1=1,A_2=1,\eta_{11}=1)$ and $(A_1=1\oplus\psi,A_2=1,\eta_{11}=\eta_{\psi 1}=\frac{1}{\sqrt{2}})$ characterize a GDW that is composed of the GB of the doubled Ising and the $m$-condensed boundary of the toric code. 
    
    \item $(A_1=1,A_2=1\oplus\psi,\eta_{11}=\eta_{1 \psi}=\frac{1}{\sqrt{2}})$ and $(A_1=1\oplus\psi,A_2=1\oplus\psi,\eta_{11}=\eta_{1\psi} = \eta_{\psi 1} = \eta_{\psi\psi}=\frac{1}{2})$ correspond to a GDW that is composed of the GB of the doubled Ising and the $e$-condensed boundary of the toric code. 
    
    \item $(A_1=1,A_2=1\oplus\psi,\eta_{11}=\eta_{\psi\psi}=\frac{1}{\sqrt{2}})$ characterizes the $\psi\bar{\psi}$-condensing GDW in the Ising LW model. 

\end{enumerate}

\section{GDW-GB Correspondence \label{section: folding trick}}
\begin{figure}[htbp]
    \centering
    \input{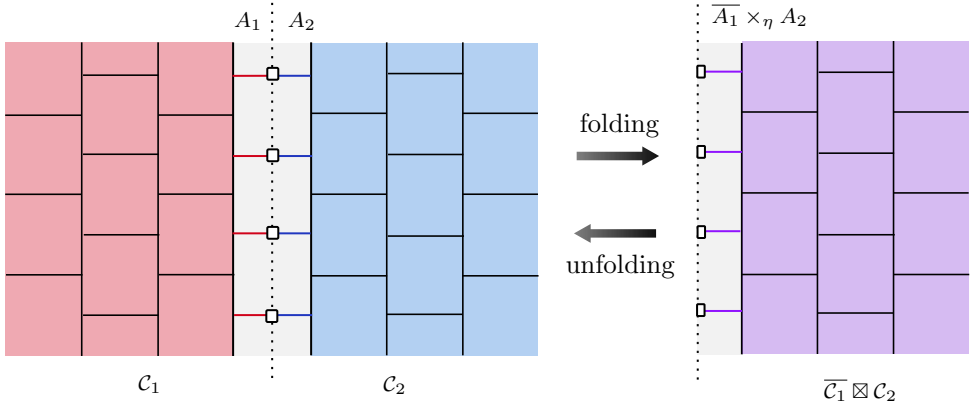}
    \caption{Fold the lattice for the ground states along the middle (dotted) line of the GDW. Then the DOFs at the gluing points in the GDW becomes those at the endpoints of tails (colored purple) in the GB. Since the DOFs at the endpoints of the tails are fixed by $M_0$, we can omit it when discussing about the ground states. }
    \label{fig: folding trick}
\end{figure}

Seen in Fig. \ref{fig: folding trick}, our lattice is symmetric with respect to the middle line in the GDW, thus is convenient for investigating the folded phase on the lattice. After folding our model along the middle line, we find that the folded phase is described by the LW model with input UFC $\mathcal{C}_1^\text{op} \boxtimes \mathcal{C}_2$, and the GDW would become a boundary of the folded phase. Here, $\boxtimes$ denotes the Deligne's tensor product, and $\mathcal{C}_2^\text{op}$ is the opposite category of $\mathcal{C}_2$~\cite{fuchs2013bicategories}.
In this section, we show that each triple $(A_1,A_2,\eta)$ characterizing a GDW would correspond to a Frobenius algebra in the input $\mathcal{C}_1^\text{op} \boxtimes \mathcal{C}_2$ for the folded phase.

We pair edges in $A_1$ to edges in $A_2$ of the GDW if they overlap after folding, and use these pairs to label the open edges (colored purle) at the boundary of the folded phase. 
Now, we can show that the boundary derived by folding is a GB of the folded phase, which is characterized by a Frobenius algebra. 
The label set for the purple edges is $L_{\eta}=\{(i,j)|\eta_{ij}\neq 0, i\in A_1, j\in A_2\}$ defined in Eq. \eqref{eq: L_eta}, which forms an algebra with the multiplication defined by:
\begin{equation}
    \Omega_{(i^\prime,i)(j^\prime,j)(k^\prime, k)} \triangleq \bar{f}_{i^\prime j^\prime k^\prime} g_{ijk} \Delta[\eta_{i^\prime i} \eta_{j^\prime j} \eta_{k^\prime k}] = \bar{f}_{i^\prime j^\prime k^\prime} g_{ijk}, 
    \label{eq: multiplication of bdry algebra of folded phase}
\end{equation}
where $\bar{f}_{abc}=f_{bac}$. The second equality holds because $(i^\prime,i),(j^\prime,j),(k^\prime, k)\in L_\eta$. In the folded phase, the quantum dimension of an $\eta$-paired object is the product of the quantum dimensions of the two objects in the pair, i.e. $d_{(i^\prime, i)}=d_{i^\prime} d_i$. We denote the algebra by $\bar{A_1}\times_\eta A_2$, where $\times_\eta$ represents the $\eta$-pairing between $A_1$ and $A_2$. It is equipped with the following properties.

\begin{itemize}
    \item Unit condition: The unit object $(1,1)$ in $L_\eta$ is the pair of the units in $L_{A_1}$ and $L_{A_2}$. Then, due to \eqref{eq: properties of input algebra} and \eqref{eq: multiplication of bdry algebra of folded phase}, the multiplication $\Omega$ should satisfy the following unit condition
\begin{equation}
    \Omega_{(a, i)(a^*, i^*)(1,1)} = \Omega_{(a,i)(1,1)(a^*, i^*)} = \Omega_{(1,1)(a,i)(a^*, i^*)} =1,
    \label{eq: unit condition for eta-alegbra}
\end{equation}
 \item Cyclic condition: 
\begin{equation}
    \Omega_{(a, i)(b, j)(c, k)} =\Omega_{(b, j)(c, k) (a, i)}. 
    \label{eq: cyclic condition for eta-algebra}
\end{equation}
    \item Associativity condition: According to the commutativity~\eqref{eq: domain wall Bp commutativity} for $(A_1,A_2,\eta)$, the algebra should also satisfy the following equation:
\begin{equation}
    \mathcal{T} \ket[\Big]{\input{images/folding/fig__Frobenius_condition_1}} = \ket[\Big]{\input{images/folding/fig__Frobenius_condition_2}},
    \label{eq: associativity condition for eta-algebra}
\end{equation}
which is explicitly the associativity condition for the boundary Frobenius algebra defined in \cite{hu2018boundary}. 
\item Strong condition: the projective condition~\eqref{eq: domain wall Bp projective} is the same as the strong condition of the Frobenius algebra:
\begin{equation}
    \mathcal{T}\left( \sum_{(i^\prime,i),(j^\prime,j)\in L_\eta} (\eta_{i^\prime i} \eta_{j^\prime j})^2 \ket[\Big]{\input{images/folding/fig__strong_condition_1}}\right) = (\eta_{k^\prime k})^2 \ket[\Big]{\input{images/folding/fig__strong_condition_2}}. \label{eq: strong condition for eta-algebra}
\end{equation}
\end{itemize}

Equations \eqref{eq: unit condition for eta-alegbra}, \eqref{eq: cyclic condition for eta-algebra}, \eqref{eq: associativity condition for eta-algebra} and \eqref{eq: strong condition for eta-algebra} together render the boundary algebra $(L_A,\Omega)$ a Frobenius algebra. 
Moreover, the Pachner moves in the GB of the folded model can be derived from the Pachner moves in the GDW:
\begin{align}
    & \mathcal{T} \ket[\Big]{\input{images/folding/fig__pachner_move_2}} = \sum_{(j^\prime,j) ,(i^\prime,i) \in L_{\eta}} \frac{u_{i^\prime} u_i u_{j^\prime} u_j}{u_{k^\prime} u_k} \Omega_{(j^\prime j) (i^\prime i) (k^\prime k)} \ket[\Big]{\input{images/folding/fig__pachner_move_1}}, \\
    & \mathcal{T} \ket[\Big]{\input{images/folding/fig__pachner_move_1}} = \frac{u_{i^\prime} u_i u_{j^\prime} u_j}{u_{k^\prime} u_k} \Omega_{(j^\prime j) (i^\prime i) (k^\prime k)} \ket[\Big]{\input{images/folding/fig__pachner_move_2}}, 
\end{align}
which align with the boundary Pachner moves derived in \cite{hu2018boundary}. 

\vspace{2ex}
\textbf{Example:} Let us consider the 6 GDWs in the $\mathbb{Z}_2$ toric code phase. After folding, these GDWs become the GBs of a 2-layered toric code phase. 
Type 1-Type 6 GDWs would produce Frobenius algebras with the following multiplications, which have cyclic symmetry:
\begin{itemize}
    \item $\text{DW}_1$: $L_\eta=\{(1,1)\}$
    \begin{equation}
        \Omega_{(1,1)(1,1)(1,1)}=1
        \label{eq: Multipilication after Folding-type1}
    \end{equation}
    \item $\text{DW}_2$: $L_\eta=\{(1,1),(1,\psi)\}$
    \begin{equation}
        \Omega_{(1,1)(1,1)(1,1)}=\Omega_{(1,1)(1,\psi)(1,\psi)}=1
        \label{eq: Multipilication after Folding-type2}
    \end{equation}
    \item $\text{DW}_3$: $L_\eta=\{(1,1),(\psi,1)\}$
    \begin{equation}
        \Omega_{(1,1)(1,1)(1,1)}=\Omega_{(\psi,1)(\psi,1)(1,1)}=1
        \label{eq: Multipilication after Folding-type3}
    \end{equation}
    \item $\text{DW}_4$: $L_\eta=\{(1,1),(1,\psi),(\psi,1),(\psi,\psi)\}$
    \begin{equation}
        \begin{aligned}
    &\Omega_{(1,1)(1,1)(1,1)}=\Omega_{(1,1)(1,\psi)(1,\psi)}=\Omega_{(1,1)(\psi,1)(\psi,1)}=1, \\ & \Omega_{(1,1)(\psi,\psi)(\psi,\psi)}=\Omega_{(1,\psi)(\psi,\psi)(\psi,1)} = \Omega_{(1,\psi)(\psi,1)(\psi,\psi)}=1. \label{eq: Multipilication after Folding-type4}
\end{aligned}
    \end{equation}
    \item $\text{DW}_5$: $L_\eta=\{(1,1),(\psi,\psi)\}$
    \begin{equation}
        \Omega_{(1,1)(1,1)(1,1)}=\Omega_{(1,1)(\psi,\psi)(\psi,\psi)}=1. \label{eq: Multipilication after Folding-type5}
    \end{equation}
    \item $\text{DW}_6$: For $e-m$ exchanging GDW, we have transformed the basis of the right model. So the label set of algebra $\bar{A_1}\times A_2$ transforms from $L_\eta=\{(1,M_0),(\psi,M_1)\}$ to $L_\eta=\{(1,1),(1,\psi),(\psi,1),(\psi,\psi)\}$. Eq. \eqref{eq: multiplication of bdry algebra of folded phase} is modified as follows:
\begin{equation}
    \Omega_{(a_1.b_1),(a_2,b_2),(a_3,b_3)}= \sum_{M_i,M_j,M_k} \bar{f}_{a_1 a_2 a_3} \Tilde{g}^{M_i M_j M_k}_{b_1 b_2 b_3} \,\ \Delta(\eta_{a_1 M_i} \eta_{a_2 M_j} \eta_{a_3 M_k}),
    \label{eq: modified multiplication of bdry algebra}
\end{equation}
where $\Tilde{g}^{M_i M_j M_k}_{b_1 b_2 b_3}$ are the expansion coefficients for the multiplication of $A_2$, which is given in Appendix .\ref{Appendix: A1-bimodules-Z2}. Then,
\begin{equation}
\begin{aligned}
    &\Omega_{(1,1)(1,1)(1,1)}=\Omega_{(1,1)(1,\psi)(1,\psi)}=1, \\ &\Omega_{(1,1)(\psi,1)(\psi,1)}=1, \Omega_{(1,1)(\psi,\psi)(\psi,\psi)}=1, \\ &  \Omega_{(1,\psi)(\psi,\psi)(\psi,1)}=i, \  \Omega_{(1,\psi)(\psi,1)(\psi,\psi)}=-i.
\end{aligned}
\end{equation}
\end{itemize}

\newpage

\section{Conclusions and Outlook}
In this paper, we have developed a novel framework for constructing GDWs within the LW model. By sewing two LW models along their open sides using algebra objects from their respective input UFCs, we provide a systematic approach that results in exactly solvable, gapped Hamiltonians to describe all GDWs between the two models. The construction introduces new DOFs at the joining point, which are captured by $\AAeta$-bimodules. This approach offers a unified description of GDWs, including those that involve anyon condensation and $e$-$m$ exchange.

Our framework for constructing GDWs is fully compatible with the folding trick. Specifically, after folding the lattice along the GDW, the algebra $\bar{A_1} \times_\eta A_2$ emerges as the Frobenius algebra that characterizes the GBs of the folded model with input ${\mathcal{C}_1}^\text{op} \boxtimes \mathcal{C}_2$, where $\times_\eta$ denotes the $\eta$-pairing between $A_1$ and $A_2$. This connection between GDWs and GBs highlights the robustness and topological nature of the construction.

While our approach provides a novel and general method for constructing GDWs, there are several avenues for future work along this line:

\begin{itemize}
    \item[(1)] Our method is able to classify GDWs by their Hamiltonians, which is at the level of input UFC. 
    Nevertheless, there exists an important subtlety in the classification, arising between Morita equivalence and physical equivalence: the Morita equivalent GBs or GDWs are not necessarily physical equivalent. One featuring example is the doubled Ising, whose GBs are characterized by two distinct but Morita-equivalent Frobenius algebras: one is $A=1$ (smooth boundary), the other is $A=1\oplus \psi$ (rough boundary). They correspond to GB with charge excitations and a GB with charge-flux composite excitations, respectively. So these two gapped boundaries are physically distinct \cite{cheng2023precision, chen2024cft}. Nevertheless, it remains an question to us whether this correspondence holds for all cases.

    \item[(2)] The fusion of GDWs and the construction of junctions between them, which are codimension-2 defects, is crucial for the completeness of the extended TQFT. For example, consider a GDW that separates the doubled Ising model and the toric code, and another GDW that realizes $e$-$m$ exchange in the toric code. When these two GDWs fuse, a junction appears where they meet. A detailed understanding of how to construct such junctions and the corresponding excitation spectrum at the junctions is a key open problem. Developing a framework to describe these junctions will provide a more complete picture of topological defects.
\end{itemize}

In conclusion, our work provides a systematic and concrete approach to studying GDWs in topologically ordered phases, solely within the bulk degrees of freedom from input UFCs. Future research will aim to extend this framework to higher dimensions and explore the full range of physical phenomena that can emerge from the fusion and interaction of these GDWs.

\acknowledgments
YC thank Yingcheng Li, Hongyu Wang and Zhihao Zhang for helpful discussions. YW is supported by NSFC Grant No. KRH1512711, the Shanghai Municipal Science and Technology Major Project (Grant No. 2019SHZDZX01), Science and Technology Commission of Shanghai Municipality (Grant No. 24LZ1400100), and the Innovation Program for Quantum Science and Technology (No. 2024ZD0300101). YW is grateful for the hospitality of the Perimeter Institute during his visit, where part of this work is done. This research was supported in part by the Perimeter Institute for Theoretical Physics. Research at Perimeter Institute is supported by the Government of Canada through the Department of Innovation, Science and Economic Development and by the Province of Ontario through the Ministry of Research, Innovation and Science. YH is supported by NSFC (Grant No. 12375001), National Key Research and Development Program of China (Grant No. 2024YFA1408900), and Zhejiang Provincial Natural Science Foundation of China (Grant No. LY23A050001).

\newpage

\appendix

\section{Review of the LW model \label{section:Review of the LW model}}
\subsection{The original LW model}

The LW model, also called the string net model \cite{levin2005string}, is described by an input unitary fusion category $\mathcal{C}$ with data $\{d_j,N_{ij}^k, G_{ijk}^{mnl}\}$. It is defined on a two-dimensional trivalent lattice with oriented edges denoted by simple objects of the input UFC $\mathcal{C}$. Every object has its dual object, which can be obtained by reversing the orientation of an edge. Denote the label set of all simple objects in $\mathcal{C}$ by $L$. The fusion rule of strings is defined by $i\otimes j=\oplus_k N_{ij}^k k$, where $N: L\times L\times L \rightarrow \mathbb{N}$. 
The fusion coefficient $N_{ij}^k$ should satisfy the following conditions that for any $a,b,c,d\in L$, 
\begin{align}
    N_{0a}^b&=N_{a0}^b=\delta_{ab},\\
    N^0_{ab}&=\delta_{ab^*},\\
\sum_{x\in L}N^x_{ab}N^d_{xc}&=\sum_{x\in L}N^d_{ax}N^x_{cb}.
\end{align}
Every three edges meeting at the same vertex should obey the fusion rule in a proper lattice configuration. When the category is multiplicity-free, $N_{ij}^k=\delta_{ijk^*}$. 

The quantum dimension $d_j$ of string type $j$ is the one-dimensional representation of the fusion rule, which satisfies:
\begin{equation}
    d_i d_j=\sum_k N_{ij}^k d_k.
\end{equation}

Finally, the unitary symmetric tetrahedral $6j$ symbols are denoted by $G_{ijk}^{lnm}$. They satisfy the following conditions:
\begin{align}
&\text{tetrahedral symmetry: }G^{ijm}_{kln}=G^{mij}_{nk^*l^*}=G^{klm^*}_{ijn^*}=\alpha_m\alpha_n\bar{G^{j^*i^*m^*}_{l^*k^*n}},  \label{eq: 6j symmetry condition}\\ &\text{associativity: } \sum_n \mathrm{d}_n G^{mlq}_{kp^*n}G^{jip}_{mns^*}G^{js^*n}_{lkr^*}=G^{jip}_{q^*kr^*}G^{riq^*}_{mls^*}, \label{eq: 6j associativity condition} \\ &\text{orthogonality: }
\sum_n \mathrm{d}_n G^{mlq}_{kp^*n}G^{l^*m^*i^*}_{pk^*n}=\frac{\delta_{iq}}{\mathrm{d}_i}\delta_{mlq}\delta_{k^*ip}. \label{eq: 6j orthogonal condition}
\end{align}

The Hamiltonian of the bulk LW model is the sum of all vertex operators $A_v$ and plaquette operators $B_p$, which are commutative projectors: 
\begin{equation}
    H=-\sum_v A_v-\sum_p B_p. 
\end{equation}
The vertex operator $A_v$ is to check if the trivalent vertex $v$ satisfies the fusion rule. It indicates a charge excitation at vertex $v$ if $\delta_{ijk^*}=0$, which is a discrete version of the Gauss law for electric fields. 
\begin{equation}
    A_v\ket[\Big]{\input{images/Review/fig__traditional_Av}}=\delta_{ijk^*}\ket[\Big]{\input{images/Review/fig__traditional_Av}}.
\end{equation}
The $B_p$ is a plaquette operator defined by the following equation, which means inserting string loops of all types into plaquette $p$ and fusing them with the boundary of the plaquette.
\begin{equation}
    B_p=\frac{1}{D}\sum_s d_s B_p^s,
\end{equation}
where $D$ is the total quantum dimension $D=\sum_{s\in \mathrm{Ob}(\mathcal{C})} d_s^2$ and 
\begin{align}
    B_p^s \ket[\Big]{\input{images/Review/fig__traditional_Bp_2}}&=\ket[\Big]{\input{images/Review/fig__traditional_Bp_3}}
    =\sum_{i_1,i_2,i_3,i_4,i_5,i_6,s\in \mathcal{C}}G_{s^*i_6^*i_1}^{k_1^*i_1^{\prime}{i_6^{\prime}}^*} G_{i_2s i_1}^{{i_1^{\prime}}^*k_2i_2^{\prime}}\\G_{i_3^*si_2}^{{i_2^{\prime}}^*k_3 {i_3^{\prime}}^*}G_{s^*i_3i_4^*}^{k_4{i_4^{\prime}}^*i_3^{\prime}}G_{s^*i_4i_5^*}^{k_5^*{i_5^{\prime}}^*i_4^{\prime}}G_{i_6 s i_5^*}^{i_5^{\prime}k_6^*i_6^{\prime}}&v_{i_1}v_{i_2}v_{i_3}v_{i_4}v_{i_5}v_{i_6}v_{i_1^{\prime}}v_{i_2^{\prime}}v_{i_3^{\prime}}v_{i_4^{\prime}}v_{i_5^{\prime}}v_{i_6^{\prime}}\ket[\Big]{\input{images/Review/fig__traditional_Bp_1}}
\end{align}

\subsection{The tailed LW model with enlarged Hilbert space \label{Appendix: the extended LW}}

\begin{figure}[htbp]
    \centering
    \input{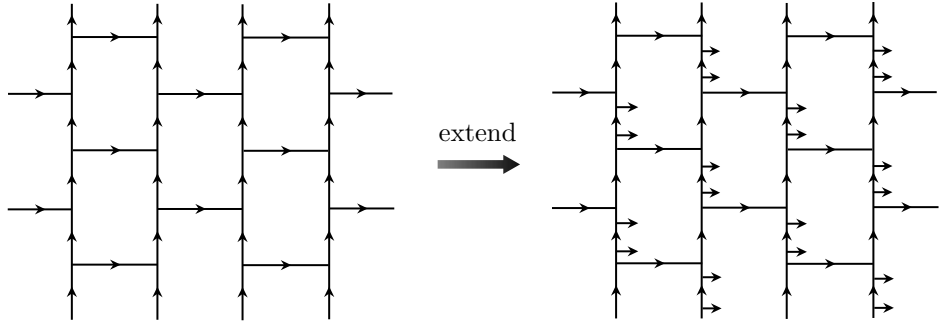}
    \caption{The extended LW model by enlarging the Hilbert space}
    \label{fig:the extended LW}
\end{figure}

The tailed LW model is an extension of the original LW model, and is useful for investigating quasiparticle excitations. The Hilbert space of the conventional LW model is enlarged to hold charge excitations. As shown in Fig. \ref{fig:the extended LW}, we associate a tail to each vertex in the conventional LW model and the DOFs of those tails take value in $\mathcal{C}$. 
To distinguish the vertices emerging after introducing tails from the vertices in the conventional model, we call the original vertices the primary vertices, and call the new vertex joining one tail and two bulk edges the secondary vertex the secondary vertex. Then, we find that the tails can present the internal charge DOFs of quasiparticle excitations to every primary vertex~\cite{hu2018full}. Now, the Hilbert space is spanned by configurations of all the edge DOFs on the lattice, including the tails. At each trivalent vertex for the basis states in the Hilbert space, the fusion rules should always be satisfied.

The Hamiltonian of the tailed LW model is also the sum of all vertex operators and plaquette operators:
\begin{equation}
    H=-\sum_v A_v - \sum_p B_p.
\end{equation}
The vertex operator defined for each primary vertex is modified as follows:
\begin{equation}
    \mathcal{A}_v \ket[\Big]{\input{images/appendix/fig-Extended_LW-Av}} = \delta_{q,1} \ket[\Big]{\input{images/appendix/fig-Extended_LW-Av}}.
\end{equation}
Since the fusion rules are always satisfied, we have $\delta_{ijk_1}=1,\delta_{k_1 k_2 q}=1$ for the basis state above. One can see that the action of the vertex operators is to project a state to the ground state without any excitations at the vertices.
The plaquette operator is
\begin{equation}
    B_p =\frac{1}{D} \sum_s d_s B_p^s,
\end{equation}
with
\begin{equation}
\begin{aligned}
    B_p^s \ket[\Big]{\input{images/appendix/fig-Extended_LW-Bp1}} = \delta_{q_1,1} \delta_{q_2,1} \sum_{j_1,j_2,i_3,i_4,i_5,i_6,i_7,i_8} G^{s^* i_8^* i_8^\prime}_{k_1^* j_1^\prime j_1^*} G^{s^* j_1^* j_1^\prime}_{k_2 i_3^\prime i_3^*} G^{s^* i_3^* i_3^\prime}_{k_3 {i_4^\prime}^* i_4} & \\
    \times G^{s^* i_4 {i_4^\prime}^*}_{k_4 {i_5^\prime}^* i_5} G^{s^* i_5 {i_5^\prime}^*}_{q_4^* {j_2^\prime}^* i_2} G^{s^* j_2 {j_2^\prime}^*}_{q_3^* {i_6^\prime}^* i_6} G^{s^* i_6 {i_6^\prime}^*}_{k_5^* {i_7^\prime}^* i_7} G^{s^* i_7 {i_7^\prime}^*}_{k_6^* i_8^\prime i_8^*}  v_{j_1} v_{j_1^\prime} v_{j_2} v_{j_2^\prime} v_{i_3} v_{i_3^\prime} v_{i_4} v_{i_4^\prime} v_{i_5} v_{i_5^\prime} v_{i_6} v_{i_6^\prime}& \\
    \times v_{i_7} v_{i_7^\prime} v_{i_8} v_{i_8^\prime} \ket[\Big]{\input{images/appendix/fig-Extended_LW-Bp2}}. &
\end{aligned}
\end{equation}

The plaquette operator in the extended Hamiltonian projects out the states with nontrivial tails in that plaquette. Moreover, the ground states are the eigenvectors of all plaquette operators and vertex operators that correspond to the eigenvalues +1. 
one can observe that the ground states of the tailed LW model are exactly the same as those of the traditional LW model.

\section{Ribbon operators in the bulk} \label{Appendix: ribbon operators in the bulk}
The ribbon operators to create elementary excitations are characterized by the minimal solutions of half-braiding tensors, which capture the effects of a quasiparticle moving across an edge. These tensors encode essential information such as changes in the internal charges of the quasiparticle and the phase factor associated with the braiding process~\cite{christian2023lattice, hu2018full}.
The following equation illustrates a ribbon operator acting on the neighboring plaquettes with a shared edge $E$ and creates a pair of dyons $(J,p),(J,q)$ at its ends:
\begin{equation}
    W^{J;pq}_E \ket[\Big]{\input{images/bulk_ribbon/ribbon_operator_1}}=\sum_{l^\prime} \frac{v_{l^\prime}}{v_l} z^J_{pl^\prime q l} \ket[\Big]{\input{images/bulk_ribbon/ribbon_operator_2}} = \ket[\Big]{\input{images/bulk_ribbon/ribbon_operator_3}},
\end{equation}
where $z^J$ is the half-braiding tensor of the quasiparticle (dyon species) $J$. The half-braiding tensor satisfies the following naturality condition. 
\begin{equation}
    \ket[\Big]{\input{images/bulk_ribbon/half-braiding_tensor_1}}=\ket[\Big]{\input{images/bulk_ribbon/half-braiding_tensor_2}}.
    \label{eq: half-braiding tensor formula}
\end{equation}
It is formulated as
\begin{equation}
    \sum_{k^\prime ,i^\prime ,j^\prime ,j_1}d_{k^\prime }d_{i^\prime } z^J_{pk^\prime qk} z^J_{q^*i^\prime ri} G^{k^\prime pk}_{jij^\prime } G^{j^\prime ik^\prime }_{qki^\prime } G^{rii^\prime }_{kj^\prime j_1} v_{j_1}=\sum_{j^\prime }\frac{1}{v_{j}} z^J_{pj^\prime rj}.
    \label{eq: naturality condition of z}
\end{equation}

The measuring operators for a dyon species $J$ is defined as
\begin{equation}
    \Pi_J=\sum_{q}\Pi^J_q, \quad \Pi^J_q=\sum_{s,t}\Pi^J_{qst} B_{qsqt}, \label{eq: bulk measuring operator}
\end{equation}
where $q$ is the internal charge of $J$ and 
\begin{equation}
    B_{qsqt} \ket[\Big]{\input{images/bulk_ribbon/measuring_operator_1}}=\ket[\Big]{\input{images/bulk_ribbon/measuring_operator_2}}.
\end{equation}
When $J=J_0$ is a trivial excitation of zero charge, the corresponding measuring operator is $\Pi_{J_0}=\sum_s \Pi^{J_0}_{1ss} B_{1s1s}$. Here, the coefficient $\Pi^{J_0}_{1ss}=d_s/D$ and $B_{1s1s}$ is the same as the plaquette operator $B_p$ in the traditional LW model. Also, not only $\Pi_q^J$ for each $(J,q)$ is a projector, but their sum $\Pi_J$ is a projector because of the orthogonality condition $\Pi^J_q \cdot \Pi^J_{q^\prime}=\delta_{q q^\prime} \Pi^J_q$. Since $\Pi_J$ is spanned by the basis $B_{qsqt}$, we can compute the following equation
\begin{equation}
    \sum_{m,n,s,t,l,r} \Pi^J_{qns} \Pi^J_{qmt} \ket[\Big]{\input{images/Review/fig_measuring_op_in_the_bulk_1}} = \sum_{l,r} \Pi_{qlr} \ket[\Big]{\input{images/Review/fig_measuring_op_in_the_bulk_2}},
    \label{eq: tube algebra}
\end{equation}
and then derive the orthogonality condition for $\Pi^J_{qst}$:
\begin{equation}
    d_l d_r \sum_{m,n,s,t} \Pi^J_{qns} \Pi^J_{qmt} G^{m^* t q^*}_{n s^* r} G^{m^*rs^*}_{q^*n^*l}G^{t^*qm}_{ln^*r}
    =\Pi^J_{qlr}.
    \label{eq: bulk measuring op-coefficient}
\end{equation}

Also from Eq. \ref{eq: tube algebra}, we can see these measuring operators form a tube algebra and we denote it by $\mathcal{A}$. Since each $\Pi^J$ correspond to a quasiparticle $J$, the states with elementary excitations are characterized by irreducible representations of $\mathcal{A}$. Additionally, $\Pi^J_q$ living in a subspace spanned by $B_{qsqt}$ with fixed charge $q$ can form a subalgebra $\mathcal{A}_q$. Dyons are identified by operators $\Pi^J_q$, each of which corresponding to the minimal solutions of Eq. \ref{eq: bulk measuring op-coefficient}. The minimal solution for means that if $\Pi^J_q = ({\Pi^J_q})_1 + ({\Pi^J_q})_2$, then either $({\Pi^J_q})_1$ or $({\Pi^J_q})_2$ is zero.

Since the state $W^{J;pq}\ket[\Big]{\psi_{gs}}$ with dyon $(J,q)$ excitations is the $+1$ eigenstate of $\Pi^J_q$, the coeffients $\Pi^J_{qst}$ shows close relation with the half-braiding tensors $z^J_{qsqt}$:
\begin{equation}
    \frac{\Pi^J_{qst}}{\Pi^J_{q1q}}=\frac{d_s d_t}{d_q} z^J_{qsqt}.
\end{equation}

We present several examples in the following sections.

\subsection{The LW \texorpdfstring{$\mathbb{Z}_2$ with input $\mathcal{R}ep(\mathbb{Z}_2)$}{}} \label{Appendix: Half-braiding tensors-Z2}
There are four dyon species $1,e,m,\varepsilon$ in the $\mathbb{Z}_2$ LW model, and the minimal solutions have a one-to-one correspondence with them as follows:

 {\large $1$}
\begin{equation}
    z_{1111} 
    = 1, \quad z_{1\psi 1\psi} = 1 \label{eq: RepZ2-trivial anyon}
\end{equation}

{\large $e$}
\begin{equation}
    z_{\psi\psi\psi1} = 1, \quad z_{\psi1\psi\psi}     =1 \label{eq: RepZ2-e}
\end{equation}

{\large$m$}
\begin{equation}
    z_{1111} 
    = 1,\quad z_{1\psi1\psi} = -1 \label{eq: RepZ2-m}
\end{equation}

{\large$\varepsilon$}
\begin{equation}
   z_{\psi1\psi\psi} = -1,\quad z_{\psi\psi\psi1}     
    = 1 \label{eq: RepZ2-epsilon}
\end{equation}

\subsection{The LW \texorpdfstring{$\mathbb{Z}_2$ with input $\mathcal{V}ec(\mathbb{Z}_2)$}{}} \label{Appendix: Half-braiding tensors2-Z2}

As mentioned in Section \ref{sec: e-m exchanging GDW in TC}, $\mathcal{V}ec(\mathbb{Z}_2)$ has two simple objects represented by the irreducible $A$-$A$-bimodules $N_0$ and $N_1$ of $\mathcal{R}ep(\mathbb{Z}_2)$, where $A=1\oplus\psi$ the nontrivial Frobenius algebra in $\mathcal{R}ep(\mathbb{Z}_2)$. The minimal solutions to Eq. \eqref{eq: naturality condition of z} for $\mathcal{V}ec(\mathbb{Z}_2)$ are given below:

{\large $1$}
\begin{equation}
    z_{N_0 N_0 N_0 N_0} 
    = 1, \quad z_{N_0 N_1 N_0 N_1} = 1 \label{eq: VecZ2-trivial anyon}
\end{equation}

{\large $m$}
\begin{equation}
    z_{N_1 N_1 N_1 N_0} = 1, \quad z_{N_1 N_0 N_1 N_1} = 1 \label{eq: VecZ2-m}
\end{equation}

{\large$e$}
\begin{equation}
    z_{N_0 N_0 N_0 N_0} 
    = 1,\quad z_{N_0 N_1 N_0 N_1} = -1 \label{eq: VecZ2-e}
\end{equation}

{\large$\varepsilon$}
\begin{equation}
   z_{N_1 N_0 N_1 N_1 } = -1,\quad z_{N_1 N_1 N_1 N_0}     
    = 1 \label{eq: VecZ2-epsilon}
\end{equation}
The categories $\mathcal{V}ec(\mathbb{Z}_2)$ and $\mathcal{R}ep(\mathbb{Z}_2)$ are equivalent. One can observe that replacing $N_0,N_1$ of $\mathcal{V}ec(\mathbb{Z}_2)$ with $1,\psi$, respectively would reproduce the fusion rules, 6j-symbols of $\mathcal{R}ep(\mathbb{Z}_2)$. 
As a result, solutions to Eq. \eqref{eq: half-braiding tensor formula} of half-braiding tensors in the basis $\{N_0,N_1\}$ have one-to-one correpondence with those in the basis $\{1,\psi\}$. Nevertheless, the correspondence between the minimal solutions of half-braiding tensors and the species of quasiparticles in the $\mathcal{V}ec(\mathbb{Z}_2)$-model differs from that in the $\mathcal{R}ep(\mathbb{Z}_2)$-model: $e$ and $m$ in Eq. \eqref{eq: RepZ2-e} and \eqref{eq: RepZ2-m} are exchanged in Eq. \eqref{eq: VecZ2-e} and \eqref{eq: VecZ2-m}.  

The exchange of $e$ and $m$ can be understood through the symmetry transformations. Any basis state of the $\mathcal{V}ec(\mathbb{Z}_2)$-model transforms into a basis state of the $\mathcal{R}ep(\mathbb{Z}_2)$ as follows: (1) Each edge labeled by $N_0$ or $N_1$ is mapped to $\{1,\psi\}$; (2) Each vertex in the bulk satisfies the following expansion \cite{zhao2024noninvertible}:

\begin{equation}
    \tikzset{every picture/.style={line width=0.75pt}} 
\begin{tikzpicture}[x=0.75pt,y=0.75pt,yscale=-1,xscale=1, baseline=(XXXX.south) ]
\path (0,57);\path (75.31640625,0);\draw    ($(current bounding box.center)+(0,0.3em)$) node [anchor=south] (XXXX) {};
\draw [color={rgb, 255:red, 33; green, 55; blue, 191 }  ,draw opacity=1 ][line width=0.75]    (56.75,52.52) -- (34.71,30.48) ;
\draw [shift={(47.78,43.55)}, rotate = 225] [fill={rgb, 255:red, 33; green, 55; blue, 191 }  ,fill opacity=1 ][line width=0.08]  [draw opacity=0] (5.36,-2.57) -- (0,0) -- (5.36,2.57) -- (3.56,0) -- cycle    ;
\draw [color={rgb, 255:red, 33; green, 55; blue, 191 }  ,draw opacity=1 ][line width=0.75]    (12.74,52.45) -- (34.71,30.48) ;
\draw [shift={(21.67,43.52)}, rotate = 315] [fill={rgb, 255:red, 33; green, 55; blue, 191 }  ,fill opacity=1 ][line width=0.08]  [draw opacity=0] (5.36,-2.57) -- (0,0) -- (5.36,2.57) -- (3.56,0) -- cycle    ;
\draw [color={rgb, 255:red, 33; green, 55; blue, 191 }  ,draw opacity=1 ][line width=0.75]    (34.71,3.71) -- (34.71,30.48) ;
\draw [shift={(34.71,14.19)}, rotate = 90] [fill={rgb, 255:red, 33; green, 55; blue, 191 }  ,fill opacity=1 ][line width=0.08]  [draw opacity=0] (5.36,-2.57) -- (0,0) -- (5.36,2.57) -- (3.56,0) -- cycle    ;
\draw (-0.5,33.2) node [anchor=north west][inner sep=0.75pt]  [font=\footnotesize]  {$M_{k}$};
\draw (48.82,32.99) node [anchor=north west][inner sep=0.75pt]  [font=\footnotesize]  {$M_{j}$};
\draw (36.71,5.41) node [anchor=north west][inner sep=0.75pt]  [font=\footnotesize]  {$M_{i}$};
\draw (20.5,20.2) node [anchor=north west][inner sep=0.75pt]  [font=\footnotesize]  {$v$};
\end{tikzpicture}
  = \sum_{a\in L_{N_0}, b\in L_{N_0}, c\in L_{N_0}} \mathcal{V}_{abc}^{N_i N_i N_k} \tikzset{every picture/.style={line width=0.75pt}} 
\begin{tikzpicture}[x=0.75pt,y=0.75pt,yscale=-1,xscale=1, baseline=(XXXX.south) ]
\path (0,57);\path (75.31640625,0);\draw    ($(current bounding box.center)+(0,0.3em)$) node [anchor=south] (XXXX) {};
\draw [color={rgb, 255:red, 33; green, 55; blue, 191 }  ,draw opacity=1 ][line width=0.75]    (56.75,52.52) -- (34.71,30.48) ;
\draw [shift={(47.78,43.55)}, rotate = 225] [fill={rgb, 255:red, 33; green, 55; blue, 191 }  ,fill opacity=1 ][line width=0.08]  [draw opacity=0] (5.36,-2.57) -- (0,0) -- (5.36,2.57) -- (3.56,0) -- cycle    ;
\draw [color={rgb, 255:red, 33; green, 55; blue, 191 }  ,draw opacity=1 ][line width=0.75]    (12.74,52.45) -- (34.71,30.48) ;
\draw [shift={(21.67,43.52)}, rotate = 315] [fill={rgb, 255:red, 33; green, 55; blue, 191 }  ,fill opacity=1 ][line width=0.08]  [draw opacity=0] (5.36,-2.57) -- (0,0) -- (5.36,2.57) -- (3.56,0) -- cycle    ;
\draw [color={rgb, 255:red, 33; green, 55; blue, 191 }  ,draw opacity=1 ][line width=0.75]    (34.71,3.71) -- (34.71,30.48) ;
\draw [shift={(34.71,14.19)}, rotate = 90] [fill={rgb, 255:red, 33; green, 55; blue, 191 }  ,fill opacity=1 ][line width=0.08]  [draw opacity=0] (5.36,-2.57) -- (0,0) -- (5.36,2.57) -- (3.56,0) -- cycle    ;
\draw (5.5,32.7) node [anchor=north west][inner sep=0.75pt]  [font=\footnotesize]  {$c$};
\draw (48.82,33.49) node [anchor=north west][inner sep=0.75pt]  [font=\footnotesize]  {$b$};
\draw (36.32,6.49) node [anchor=north west][inner sep=0.75pt]  [font=\footnotesize]  {$a$};
\draw (20,21.7) node [anchor=north west][inner sep=0.75pt]  [font=\footnotesize]  {$v$};
\end{tikzpicture}
,
    \label{eq: basis tranformation}
\end{equation}
where the expansion coefficients are 
\begin{equation}
    \begin{aligned}
        &\mathcal{V}_{111}^{N_0 N_0 N_0}=\mathcal{V}_{1\psi\psi}^{N_0 N_0 N_0} = \mathcal{V}_{\psi 1 \psi}^{N_0 N_0 N_0} = \mathcal{V}_{\psi\psi 1}^{N_0 N_0 N_0}=1 \\
        & \mathcal{V}_{111}^{N_0 N_1 N_1}=\mathcal{V}_{1\psi\psi}^{N_0 N_1 N_1} = 1, \ \mathcal{V}_{\psi 1 \psi}^{N_0 N_1 N_1} = i, \ \mathcal{V}_{\psi\psi 1}^{N_0 N_1 N_1}=-i.
    \end{aligned}
\end{equation}
Using the basis transformation in Eq. \eqref{eq: basis tranformation} , one can show that, for instance, the half-braiding tensors for $m$ ($e$) in \eqref{eq: VecZ2-m} (\eqref{eq: VecZ2-e}) will transform into those in \eqref{eq: RepZ2-m} (\eqref{eq: RepZ2-e}).

\subsection{The Ising LW \label{Appendix: Half-braiding tensors-Ising}}

The minimal solutions to the half-braiding tensors of doubled Ising are listed below, and they are associated with the 9 dyon species of the Ising LW. Here we set $\omega=\exp{\pi i/8}=(-1)^{1/8}$. 

{\large $1\Bar{1}$}
\begin{equation}
    z_{1111} = 1, \quad 
    z_{1\psi 1\psi} = 1, \quad 
    z_{1\sigma 1\sigma} = 1
\end{equation}

{\large $\psi\Bar{\psi}$}
\begin{equation}
    z_{1111}=1, \quad
    z_{1\psi 1\psi}=1, \quad 
    z_{1\sigma 1\sigma}=-1
\end{equation}

{\large$\sigma\Bar{\sigma}$}
\begin{equation}
    \begin{aligned}
        &  z_{1111} = 1,\quad z_{1\psi1\psi} = -1, \\
        &  z_{\psi1\psi\psi} =1,\quad z_{\psi\psi\psi1} = 1, \\
        & z_{1\sigma\psi\sigma} = \pm 1, \quad z_{\psi\sigma1\sigma} = \pm 1.
    \end{aligned}
\end{equation}

{\large$\psi \Bar{1}$}
\begin{equation}
       z_{\psi1\psi\psi} = -1, \quad z_{\psi\psi\psi1} = 1, \quad z_{\psi\sigma\psi\sigma} = i.
\end{equation}

{\large$1\Bar{\psi}$}
\begin{equation}
         z_{\psi1\psi\psi} = -1, \quad z_{\psi\psi\psi1} = 1, \quad z_{\psi\sigma\psi\sigma} = -i.
\end{equation}

{\large$\sigma\Bar{1}$}
\begin{equation}
    z_{\sigma1\sigma\sigma} = \omega, \quad 
    z_{\sigma\sigma\sigma1} = 1, \quad
    z_{\sigma\sigma\sigma\psi} = i, \quad
    z_{\sigma\psi\sigma\sigma} = -i\omega.
\end{equation}

{\large$1\Bar{\sigma}$}
\begin{equation}
    z_{\sigma1\sigma\sigma} = \omega^*, \quad 
    z_{\sigma\sigma\sigma1} = 1, \quad
    z_{\sigma\sigma\sigma\psi} = -i, \quad
    z_{\sigma\psi\sigma\sigma} = i\omega^*.
\end{equation}

{\large$\sigma\Bar{\psi}$}
\begin{equation}
    z_{\sigma1\sigma\sigma} = -\omega, \quad 
    z_{\sigma\sigma\sigma1} = 1, \quad
    z_{\sigma\sigma\sigma\psi} = i, \quad
    z_{\sigma\psi\sigma\sigma} = i\omega.
\end{equation}

{\large$\psi\Bar{\sigma}$}
\begin{equation}
    z_{\sigma1\sigma\sigma} = -\omega^*, \quad 
    z_{\sigma\sigma\sigma1} = 1, \quad
    z_{\sigma\sigma\sigma\psi} = -i, \quad
    z_{\sigma\psi\sigma\sigma} =-i\omega^*.
\end{equation}

\newpage

\section{Exactly solvable conditions to determine \texorpdfstring{$B_p^\text{DW}$}{} \label{appendix: exactly solvable conditions for Bp}}

\subsection{Projective condition}
The plaquette operator $B_p^\text{DW}$ requires to be projective, i.e. $(B_p^\text{DW})^2\stackrel{!}{=}B_p^\text{DW}$, which is computed graphically as follows. Without loss of generality, we can consider the DW plaquette operator $B_p^\text{DW}$ acting on a basis state with trivial $\AAeta$-bimodules. 
\begin{align*}
    (B_p^\text{DW})^2 &\ket[\Big]{\input{images/appendix/fig-projective_condition_1}}= \sum_{i^\prime ,j^\prime \in A_1, i,j\in A_2} v_{i^\prime }v_i v_{j^\prime } v_j (\eta_{i^\prime i} \eta_{j^\prime j})^2 \ket[\Big]{\input{images/appendix/fig-projective_condition_2}}\\
    &= \sum_{i^\prime ,j^\prime \in A_1, i,j\in A_2} v_{i^\prime }v_i v_{j^\prime } v_j (\eta_{i^\prime i}\eta_{j^\prime j})^2 \sum_{k^\prime \in A_1, k\in A_2} G^{i{i^\prime}^*1}_{{j^\prime}^* j^\prime {k^\prime}^*} v_{k^\prime } G^{i^*i1}_{j^*jk}v_k \ket[\Big]{\input{images/appendix/fig-projective_condition_3}}\\
    &= \sum_{i^\prime ,j^\prime ,k^\prime \in A_1, i,j,k\in A_2} v_{k^\prime } v_k (\eta_{i^\prime i}\eta_{j^\prime j})^2 \delta_{i^\prime j^\prime k^\prime }\delta_{ijk} \ket[\Big]{\input{images/appendix/fig-projective_condition_3}}\\
    &\overset{\eqref{eq: DW Pachner move}}{=} \sum_{i^\prime ,j^\prime ,k^\prime \in A_1, i,j,k\in A_2} v_{i^\prime} v_{j^\prime} v_{i} v_{j} (\eta_{i^\prime i}\eta_{j^\prime j})^2 \delta_{i^\prime j^\prime k^\prime }\delta_{ijk} f_{{i^\prime }^*{j^\prime }^*k^\prime }f_{j^\prime i^\prime {k^\prime }^*}g_{jik^*}g_{i^*j^*k} \ket[\Big]{\input{images/appendix/fig-projective_condition_4}} \\
    &\stackrel{!}{=} \sum_{k^\prime \in A_1, k\in A_2} v_{k^\prime } v_k (\eta_{k^\prime k})^2 \delta_{i^\prime j^\prime k^\prime }\delta_{ijk} \ket[\Big]{\input{images/appendix/fig-projective_condition_4}}
    =B_p^\text{DW} \ket[\Big]{\input{images/appendix/fig-projective_condition_1}}.
\end{align*}

Therefore, the projective condition is given by
\begin{equation}
    \sum_{\substack{i^\prime ,j^\prime \in \Tilde{A}\\i,j\in A}} \sum_{\eta_{k^\prime k}\neq 0} \delta_{i^\prime j^\prime k^\prime }\delta_{ijk}\frac{v_{i^\prime }v_{j^\prime } v_i v_j}{v_k v_{k^\prime }}f_{{i^\prime }^*{j^\prime }^*k^\prime }f_{j^\prime i^\prime {k^\prime }^*}g_{jik^*}g_{i^*j^*k} (\eta_{j^\prime j}\eta_{i^\prime i})^2
    = (\eta_{k^\prime k})^2.
    \label{eq: domain wall Bp projective-appendix}
\end{equation}

It can be represented graphically as
\begin{equation}
    \mathcal{T} \Big( \sum_{i^\prime ,j^\prime \in A_1,i,j\in A_2} (\eta_{i^\prime i} \eta_{j^\prime j})^2 \ket[\Big]{\input{images/domain_wall/fig-Bp_projector_condition_1}} \Big) = \Delta[\eta_{j^\prime j} \eta_{i^\prime i} \eta_{k^\prime k}] \cdot (\eta_{k^\prime k})^2 \ket[\Big]{\input{images/domain_wall/fig-Bp_projector_condition_2}}. \label{eq: Appendix. Bp projective condition 1}
\end{equation}

\subsection{Commutative condition}

The commutativity of the plaquette operators that act on the neighboring plaquettes in the GDW means $[B_{p_1}^\text{DW},B_{p_2}^\text{DW}]=0$. Here, we compute the action of $B_{p_1}^\text{DW}B_{p_2}^\text{DW}$ on a basis state:
\begin{align*}
    &B_{p_1}^\text{DW}B_{p_2}^\text{DW}\ket[\Big]{\input{images/appendix/fig-commutativity_condition_1}}= 
   \sum_{i_1,i_2\in A_1,j_1,j_2\in A_2} v_{i_1}v_{i_2}v_{j_1}v_{j_2}(\eta_{i_1j_1}\eta_{i_2j_2})^2 \ket[\Big]{\input{images/appendix/fig-commutativity_condition_2}}\\ &= \sum_{i_1,i_2,k_2,k_3\in A_1} \sum_{j_1,j_2,l_2,l_3\in A_2}\frac{v_{k_3}v_{l_3}}{v_{k_1}v_{l_1}}(\eta_{i_1j_1}\eta_{i_2j_2})^2 \ket[\Big]{\input{images/appendix/fig-commutativity_condition_3}} \\
   &= \sum_{i_1,i_2,k_2,k_3,k_2^{\prime}\in A_1} \sum_{j_1,j_2,l_2,l_3,l_2^\prime\in A_2} \frac{v_{k_3}v_{l_3}}{v_{k_1}v_{l_1}}(\eta_{i_1j_1}\eta_{i_2j_2})^2 
 G^{k_3^*k_2i_2^*}_{i_1k_1{k_2^{\prime}}^*} G^{l_1^*j_1l_2}_{l_3j_2l_2^{\prime}} v_{l_2} v_{l_2^{\prime}} v_{k_2} v_{k_2^{\prime}} \ket[\Big]{\input{images/appendix/fig-commutativity_condition_4}}.
\end{align*}
To derive the commutativity condition for the plaquette operators, it is not necessary to fuse the entire loop. Instead, we focus on the shared edge between the two plaquettes and fuse only the relevant part of the string loop with it. Then, we compute the action of $B_{p_2}^\text{DW}B_{p_1}^\text{DW}$ and using Pachner moves to transfrom the state into the same basis:
\begin{equation*}
    B_{p_2}^\text{DW}B_{p_1}^\text{DW}\ket[\Big]{\input{images/appendix/fig-commutativity_condition_1}}= \sum_{i_1,i_2,k_2^{\prime},k_3\in A_1} \sum_{j_1,j_2,l_2^\prime,l_3\in A_2} \frac{v_{k_3}v_{l_3}}{v_{k_1}v_{l_1}}(\eta_{i_1j_1}\eta_{i_2j_2})^2 \ket[\Big]{\input{images/appendix/fig-commutativity_condition_5}}. 
\end{equation*}
Then, the coefficients before the same basis states should match, leading to the formula below: 
\begin{multline}
    \sum_{\substack{k_2,k_2^{\prime}\in A_1 \\ l_2,l_2^{\prime}\in A_2}}\sum_{\substack{\eta_{i_1 j_1},\eta_{i_2 j_2},\eta_{k_1 l_1},\\ \eta_{k_2 l_2},\eta_{k_3 l_3} \neq 0} }G^{k_3^*k_2i_2^*}_{i_1k_1{k_2^{\prime}}^*} G^{l_1^*j_1l_2}_{l_3j_2l_2^{\prime}} v_{l_2} v_{l_2^{\prime}} v_{k_2} v_{k_2^{\prime}} f_{k_2^*k_3i_2}f_{k_1^*i_1^*k_2}g_{l_2^*j_1l_1}g_{l_3^*l_2j_2^*} \\ =\sum_{\substack{k_2\prime \in A_1 \\ l_2^{\prime}\in A_1}}\sum_{\substack{\eta_{i_1 j_1},\eta_{i_2 j_2},\eta_{k_1 l_1},\\ \eta_{k_2^\prime l_2^\prime}, \eta_{k_3 l_3} \neq 0} }f_{{k_2^{\prime}}^*i_1^*k_3}f_{k_1^*k_2^\prime i_2} g_{{l_2^{\prime}}^*l_1j_2^*} g_{l_3^*j_1l_2^{\prime}}.
    \label{eq: domain wall Bp commutativity-appendix}    
\end{multline}
The formula above can be represented graphically as:
\begin{equation}
    \mathcal{T}(\ket[\Big]{\input{images/domain_wall/fig-Bp_commutativity_condition_1}})= \Delta[\eta_{i_1 j_1} \eta_{i_2 j_2} \eta_{k_1 l_1} \eta_{k_3 l_3}] \ket[\Big]{\input{images/domain_wall/fig-Bp_commutativity_condition_2}}.
    \label{eq: domain wall Bp commutativity-appendix-graphical} 
\end{equation}

\newpage

\section{Excitations in the GDW}
\subsection{Measuring operator at the GDW \label{appendix: Measuring operator at the domain wall}}

In the derivation, we use the topological invariance of the Pachner moves.

\begin{equation}
    \begin{aligned}
        \Pi_{M}  \ket[\Big]{\input{images/DW_excitations/fig__DW-measuring_operator_1}} &=  \mathcal{T} \Big( \ket[\Big]{\input{images/DW_excitations/fig__DW-measuring_operator_2}}  \xrightarrow{\eqref{eq: DW Pachner move}} \ket[\Big]{\input{images/DW_excitations/fig__DW-measuring_operator_3}} \\ & \xrightarrow{\eqref{eq: A-module}}  \ket[\Big]{\input{images/DW_excitations/fig__DW-measuring_operator_4}}  \xrightarrow{\eqref{eq: A-module}} \ket[\Big]{\input{images/DW_excitations/fig__DW-measuring_operator_5}} \\ & \xrightarrow{\eqref{eq: filled square box}} \ket[\Big]{\input{images/DW_excitations/fig__DW-measuring_operator_6}} \xrightarrow{\eqref{eq: definiton of A1-A2-etabimodule}} \delta_{M,M'} \ket[\Big]{\input{images/DW_excitations/fig__DW-measuring_operator_7}} \Big)
    \end{aligned}
\end{equation}

\subsection{Proof of \texorpdfstring{$[B_p^{\text{DW}_6}, W^{e-m}_{E_1,E_2}] = 0$}{}} \label{proof: Bp commutes with W}

First, we compute the action of $B_p^{\text{DW}_6} W_{E_1,E_2}^{e-m}$ on a basis state of a ground state:
\begin{equation}
    B_p^{\text{DW}_6} W_{E_1,E_2}^{e-m} \ket[\Big]{\input{images/appendix/fig-commutation_of_DW_Bp_and_W1}} =  B_p^{\text{DW}_6} ([z_1^e]_{\psi \bar{i_1} \psi i_1} [z_2^m]_{N_1 \bar{j_1} N_1 j_1} \input{images/appendix/fig-commutation_of_DW_Bp_and_W2}), \label{eq: commutativity1}
\end{equation}
where we use $z_1^J$ and $z_2^J$ to denote the half-braiding tensors of the $\mathcal{C}_1$-model and the $\mathcal{C}_2$-model, respectively, i.e, the $\mathcal{R}ep(\mathbb{Z}_2)$-model and the $\mathcal{V}ec(\mathbb{Z}_2)$ in the $e$-$m$ exchanging case of the toric code. According to the fusion rules of $\mathcal{R}ep(\mathbb{Z}_2)$ ($\mathcal{V}ec(\mathbb{Z}_2)$), fusing with the nontrivial object $\psi$ ($N_1$) is equivalent to flipping the spin. We denote $\bar{1}=\psi,\bar{\psi}=1$ and $\bar{N_0}=N_1, \bar{N_1}=N_0$. Thus, it is convenient to label the result of $i_1 \times \psi$ ($j_1 \times N_1$) to be $\bar{i_1}$ ($\bar{j_1}$). Then plugging $B_p^{\text{DW}_6}=\frac{B_p^{1 N_0}+B_p^{\psi N_1}}{2}$ into Eq. \eqref{eq: commutativity1}, the R.H.S. becomes: 
\begin{equation}
\begin{aligned}
    \text{R.H.S.} & = B_p^{\text{DW}_6} ([z_1^e]_{\psi \bar{i_1} \psi i_1} [z_2^m]_{N_1 \bar{j_1} N_1 j_1} \ket[\Big]{\input{images/appendix/fig-commutation_of_DW_Bp_and_W3}}) \\
    & = \frac{1}{2} [z_1^e]_{\psi \bar{i_1} \psi i_1} [z_2^m]_{N_1 \bar{j_1} N_1 j_1} (\input{images/appendix/fig-commutation_of_DW_Bp_and_W3} + \input{images/appendix/fig-commutation_of_DW_Bp_and_W4}).
\end{aligned} \label{eq: commutativity2}
\end{equation}

Second, we compute the action of $W_{E_1,E_2}^{e-m} B_p^{\text{DW}_6}$:
\begin{equation}
\begin{aligned}
     & W_{E_1,E_2}^{e-m} B_p^{\text{DW}_6} \ket[\Big]{\input{images/appendix/fig-commutation_of_DW_Bp_and_W1}} = \frac{1}{2} W_{E_1,E_2}^{e-m} (\ket[\Big]{\input{images/appendix/fig-commutation_of_DW_Bp_and_W5}}  + \ket[\Big]{\input{images/appendix/fig-commutation_of_DW_Bp_and_W6}})  \\
     &= \frac{1}{2} ([z_1^e]_{\psi \bar{i_1} \psi i_1} [z_2^m]_{N_1 \bar{j_1} N_1 j_1} \ket[\Big]{\input{images/appendix/fig-commutation_of_DW_Bp_and_W3}} + [z_1^e]_{\psi i_1 \psi \bar{i_1}} [z_2^m]_{N_1 j_1 N_1 \bar{j_1}} \ket[\Big]{\input{images/appendix/fig-commutation_of_DW_Bp_and_W4}}).
\end{aligned} \label{eq: commutativity3}
\end{equation}
According to the results of the half-braiding tensors given in \ref{Appendix: Half-braiding tensors-Z2} and \ref{Appendix: Half-braiding tensors2-Z2}, we find the ratio $[z_1^e]_{\psi i_1 \psi \bar{i_1}} / [z_1^e]_{\psi \bar{i_1} \psi i_1}=1$ for any $i_1\in \{1,\psi\}$ and $[z_2^m]_{N_1 j_1 N_1 \bar{j_1}} / [z_2^m]_{N_1 \bar{j_1} N_1 j_1}=1$ 
\vspace{1ex}
for any $j_1\in \{N_0,N_1\}$, we have $[z_1^e]_{\psi i_1 \psi \bar{i_1}} [z_2^m]_{N_1 j_1 N_1 \bar{j_1}} = [z_1^e]_{\psi \bar{i_1} \psi i_1} [z_2^m]_{N_1 \bar{j_1} N_1 j_1}$, \vspace{1ex} which implies that Eq. \eqref{eq: commutativity2} and \eqref{eq: commutativity3} yield the same result. Consequently, the DW plaquette operator $B_p^{\text{DW}_6}$
commutes with the DW-crossing operator $W^{e-m}_{E_1,E_2}$, making it the shortest ribbon operator across the GDW.

From the computation above, we find that the shortest ribbon operator across the $e$-$m$  exchanging GDW in the toric code should satisfy the two conditions: (1) the tail of 
the anyon created in the $\mathcal{R}ep(\mathbb{Z}_2)$-model and the tail of 
the anyon created in the $\mathcal{V}ec(\mathbb{Z}_2)$-model should simultaneously be 1 and $N_0$, or $\psi$ and $N_1$; (2) the ratio $[z_1^{J_1}]_{q \bar{i} q i}/[z_1^{J}]_{q i q \bar{i}}$ for any $i \in \{1,\psi\}$ and $[z_2^{J_2}]_{q \bar{j} q j}/[z_2^{J^\prime}]_{q j q \bar{j}}$ for any $j \in \{N_0, N_1\}$ should be the same,
\vspace{1ex}
so the GDW-crossing operator $W^{J_1-J_2}_{E_1,E_2}$ can commute with the plaquette operator $B_p^\text{DW}$. Taking these two conditions into account, there are only four types of GDW cross-over operator, which are $W^{1-1}_{E_1,E_2}$, $W^{e-m}_{E_1,E_2}$, $W^{m-e}_{E_1,E_2}$, $W^{\epsilon-\epsilon}_{E_1,E_2}$.

\newpage

\section{Solutions for joining functions \texorpdfstring{$\eta$}{} in specific cases \label{Appendix: specific solutions for eta}}
\subsection{Case 1: \texorpdfstring{$A_1=A_2$}{}}
Consider the case where two algebras $A_1$ and $A_2$ are the same, i.e., $L_{A_1}=L_{A_2}\subset L_1$ and $f_{abc}=g_{abc}$. For simplicity, we denote both $A_1$ and $A_2$ by $A$. Then, there is a solution for ~\eqref{eq: domain wall Bp commutativity} and~\eqref{eq: domain wall Bp projective} that $f_{ijk}=g_{ijk}=\delta_{ijk}$, and the joining function $\eta_{ij}=\sqrt{D_A^{-1}} \delta_{ij}$. Here, $\delta_{ij}$ is the Kronecker delta and $D_A=\sum_{i\in A} d_i^2$ is the total quantum dimension of the input algebra $A$ in the GDW.

We can check by putting the solution back in Eqs.~\eqref{eq: domain wall Bp commutativity} and~\eqref{eq: domain wall Bp projective}. The commutative condition~\eqref{eq: domain wall Bp commutativity} would become:
\begin{equation}
    \sum_{k_2^\prime\in A} 
    G^{k_3 i_2 k_2^*}_{k_1^* i_1^* {k_2^{\prime}}^*} G^{j_1 l_1 l_2^*}_{j_2^* l_3^* l_2^{\prime}} d_{k_2} d_{k_2^\prime} = \delta_{{k_2^{\prime}}^*i_1^*k_3} \delta_{k_1^*k_2^\prime i_2}.
    \label{eq: special solution 1-orthogonality}
\end{equation}
We can replace the $\sum_{k_2^\prime \in A}$ by $k_2^\prime \in \mathcal{C}_1$ in the LHS because the terms newly added are zero indeed. More precisely, when $k_2^\prime \notin L_A$, then $\delta_{{k_2^{\prime}} i_1 k_3^*}=0$ for $\forall i_1, k_3\in L_A$ because the multiplication of the algebra $A$ is closed, and accordingly the 6j-symbol $G^{k_3 i_2 k_2^*}_{k_1^* i_1^* {k_2^{\prime}}^*}$ would vanish. 
After rewriting the summation, Eq.~\eqref{eq: special solution 1-orthogonality} becomes exactly the orthogonality condition of 6$j$-symbols~\eqref{eq: 6j orthogonal condition} in the $\mathcal{C}_1$-model.

Moreover, the projective condition~\eqref{eq: domain wall Bp projective} becomes:
\begin{equation}
    \sum_{i,j\in A} \delta_{ijk^*} \frac{d_i d_j}{d_k} D_A^{-1} = 1, 
    \label{eq: special solution 1-projectivity}
\end{equation}

According to the fusion rules, the equation $\sum_{i\in L_1} \delta_{ijk^*} d_i= d_j d_k$ with $i,j,k\in L_1$ holds as a one-dimensional representation. Since also the multiplication $f_{ijk}=\delta_{ijk}$ is closed, we have $\sum_{i\in A} \delta_{ijk^*} d_i= d_j d_k$ for $\forall i,j,k\in L_A$. Substitute this condition into the LHS of Eq. \ref{eq: special solution 1-projectivity}, then $(\sum_{j\in A} d_j)^2 d_k D_A^{-1}=d_k$, which apparently holds. 

In particular, if $\mathcal{C}_1=\mathcal{C}_2$ and $A_1=A_2=\mathrm{Ob}(\mathcal{C}_1)$, then $\eta_{ij}=\sqrt{D^{-1}}\delta_{ij}$ corresponds to the trivial GDW, which does not differ from the bulk region. 

\subsection{Case 2: \texorpdfstring{$A_1=1$}{} or \texorpdfstring{$A_2=1$}{} }

Another case is that one of the input algebras in the GDW is trivial, say $A_2=1$ (the following analyses are the same for $A_1=1$). Then the only non-vanishing multiplication $g_{ijk}$ is $g_{111}=1$. Then the joining function in this case is $\eta_{ij}=\sqrt{d_A^{-1}} \delta_{j1}$, with $\delta_{j1}$ the Kronecker delta, and $d_{A_1}=\sum_{i\in A_1} d_i$. 

Also, plug the solution into Eq. \eqref{eq: domain wall Bp commutativity} and~\eqref{eq: domain wall Bp projective} to have a check. 
The commutative condition~\eqref{eq: domain wall Bp commutativity} would become:
\begin{equation}
    \sum_{k_2\in A_1} f_{k_2^* k_3 i_2} f_{k_1^* i_1^* k_2}\sum_{k_2^\prime\in A_1} G^{k_3^* k_2 i_2^*}_{i_1 k_1 {k_2^\prime}^*} v_{k_2} v_{k_2^\prime} = \sum_{k_2^\prime \in A_1} f_{{k_2^\prime}^* i_1^* k_3} f_{k_1^* k_2^\prime i_2},
\end{equation}
which can be expressed diagrammatically,
\begin{equation}
    \mathcal{T} \ket[\Big]{\input{images/domain_wall/fig-Frobenius_condition_1}} = \ket[\Big]{\input{images/domain_wall/fig-Frobenius_condition_2}}.
    \label{eq: Frobenius condition}
\end{equation}
This equation~\eqref{eq: Frobenius condition} is exactly the associativity condition of the input Frobenius algebras that characterize the GBs of the LW model with the input UFC $\mathcal{C}_1$ \cite{hu2018boundary}. In this case, the GDW joins one GB characterized by $A_1$ of the $\mathcal{C}_1$-model and a smooth GB with $A_2=1$ of the $\mathcal{C}_2$-model. The gluing process is trivial.

The projective condition~\eqref{eq: domain wall Bp projective} would become:
\begin{equation}
    \sum_{i^\prime,j^\prime\in A_1} \delta_{i^\prime j^\prime {k^\prime}^*} \frac{v_{i^\prime} v_{j^\prime}}{v_{k^\prime}} f_{{i^\prime}^* {j^\prime}^* k^\prime} f_{k^\prime i^\prime {j^\prime}^*} d_{A_1}=1, 
\end{equation}
This equation coincides with the strong condition of the Frobenius algebras \cite{hu2018boundary}.

\section{Minimal solutions of \texorpdfstring{$\AAeta$}{}-bimodules \label{Appendix: A1-bimodules}}

\subsection{\texorpdfstring{$\AAeta$}{}-bimodules for the LW \texorpdfstring{$\mathbb{Z}_2$}{} model \label{Appendix: A1-bimodules-Z2}}

\subsubsection{\texorpdfstring{$\mathcal{C}_1=\mathcal{C}_2=\mathcal{R}ep(\mathbb{Z}_2)$}{}}

When $A_1=A_2=1$ with $f_{111}=g_{111}=1$, and $\eta_{11}=1$, then solving Eq.~\eqref{eq: definiton of A1-A2-etabimodule} will produce the two irreducible $\AAeta$-bimodules, which are characterized by minimal solutions of $P$-tensors as below:
\begin{itemize}
    \item $M_0={1}$ 
        \begin{equation}
         P^{11}_{111}=1 
         \label{eq: AAbimoduleM0-A=1}
         \end{equation}
    \item $M_1={\psi}$ 
        \begin{equation}
        P^{11}_{\psi \psi \psi}=1
        \label{eq: AAbimoduleM1-A=1}
        \end{equation}
\end{itemize}

For $A_1=1\oplus\psi$ with $f_{abc}=\delta_{abc}$, and $A_2=1$ with $g_{111}=1$, $\eta_{ab}=\frac{1} {\sqrt{2}}\delta_{a\in \{1,\psi\}}\delta_{b,1}$, there are two irreducible $A_1\overset{\eta}{-}A_2$-bimodules:
\begin{itemize}
    \item $M_0={1\oplus\psi}$ 
        \begin{equation}
    P^{11}_{111}=1 \quad P^{11}_{\psi\psi\psi}=1 \quad 
    P^{\psi 1}_{\psi \psi 1}=1 \quad 
    P^{\psi 1}_{1 1 \psi}=1 
    \label{eq: ABbimodulesM0'}
        \end{equation}
    \item $M_1={1\oplus\psi}$ 
        \begin{equation}
        P^{11}_{111}=1 \quad P^{11}_{\psi\psi\psi}=1 \quad 
        P^{\psi 1}_{\psi \psi 1}=i \quad 
        P^{\psi 1}_{1 1 \psi}=-i
        \label{eq: ABbimodulesM1'}
        \end{equation}
\end{itemize}

For $A_1=1$ with $f_{111}=1$, and $A_2=1\oplus\psi$ with $g_{abc}=\delta_{abc}$, $\eta_{ab}=\frac{1}{\sqrt{2}}\delta_{a,1}\delta_{b\in \{1,\psi\}}$, there are two irreducible $A_1\overset{\eta}{-}A_2$-bimodules:
\begin{itemize}
    \item $M_0={1\oplus\psi}$ 
        \begin{equation}
    P^{11}_{111}=1 \quad P^{11}_{\psi\psi\psi}=1 \quad P^{1\psi}_{1\psi\psi}=1 \quad P^{1\psi}_{\psi 1 1}=1 
    \label{eq: ABbimodulesM0}
        \end{equation}
    \item $M_1={1\oplus\psi}$ 
        \begin{equation}
        P^{11}_{111}=1 \quad P^{11}_{\psi\psi\psi}=1 \quad P^{1\psi}_{1\psi\psi}=i \quad P^{1\psi}_{\psi 1 1}=-i
        \label{eq: ABbimodulesM1}
        \end{equation}
\end{itemize}

For $A_1=A_2=1\oplus \psi$ with $f_{abc}=g_{abc}=\delta_{abc}$, and $\eta_{ab}=\frac{1}{2} \delta_{a\in \{1,\psi\}} \delta_{b\in \{1,\psi\}}$, there are also two irreducible $\AAeta$-bimodules:
\begin{itemize}
    \item $M_0={1\oplus\psi}$ 
    \begin{equation}
        \begin{aligned}
            P^{11}_{111}&=1 \quad P^{11}_{\psi\psi\psi}=1 \quad P^{1\psi}_{1\psi\psi}=1 \quad P^{1\psi}_{\psi 1 1}=1 \\
            P^{\psi 1}_{\psi \psi 1}&=1 \quad 
            P^{\psi 1}_{1 1 \psi}=1 \quad P^{\psi\psi}_{1\psi 1}=1 \quad P^{\psi\psi}_{\psi 1\psi}=1 
        \end{aligned}
        \label{eq: AAbimodulesM0-A=1+psi}
    \end{equation}
    
    \item $M_1={1\oplus\psi}$ 
    \begin{equation}
        \begin{aligned}
            P^{11}_{111} &= 1 \quad P^{11}_{\psi\psi\psi} = 1 \quad P^{1\psi}_{1\psi\psi}= i \quad P^{1\psi}_{\psi 1 1} = -i \\ 
            P^{\psi 1}_{\psi \psi 1} &= i \quad 
            P^{\psi 1}_{1 1 \psi} = -i \quad
            P^{\psi\psi}_{1\psi 1} = -1 \quad P^{\psi\psi}_{\psi 1\psi} = -1
        \end{aligned}
        \label{eq: AAbimodulesM1-A=1+psi}
    \end{equation}
\end{itemize}
In this case, the $\AAeta$-bimodules are the same as the conventional $A_1$-$A_2$-bimodules.

When $\eta_{ab}=\frac{1}{\sqrt{d_{A_1} d_{A_2}}} \delta_{a\in A_1}\delta_{b\in A_2}$, the solutions above for $A_1\overset{\eta}{-}A_2$-bimodules are the same to the conventional $A_1$-$A_2$-bimodules.

For $A_1=A_2=1\oplus\psi$ with $f_{abc}=g_{abc}=\delta_{abc}$, and $\eta_{ab}= \frac{1}{\sqrt{d_A}} \delta_{ab}, \forall a,b\in A$, there are four irreducible $\AAeta$-bimodules for the $\mathbb{Z}_2$ LW model listed as follows:
\begin{itemize}
    \item $M_0=1$ 
    \begin{equation}
            P^{11}_{111}=1 \quad P^{\psi\psi}_{1\psi 1}=1 
    \end{equation}
            
    \item $M_1=\psi$
  \begin{equation}
            P^{11}_{\psi \psi\psi}=1 \quad P^{\psi\psi}_{\psi 1 \psi}=1 
    \end{equation}

    \item $M_2=1$
  \begin{equation}
            P^{11}_{111}=1 \quad P^{\psi\psi}_{1\psi 1}=-1 
    \end{equation}
    
    \item $M_3=\psi$
  \begin{equation}
            P^{11}_{\psi \psi\psi}=1 \quad P^{\psi\psi}_{\psi 1 \psi}=-1 
    \end{equation}
\end{itemize}

    \subsubsection{\texorpdfstring{$\mathcal{C}_1=\mathcal{R}ep(\mathbb{Z}_2), \mathcal{C}_2=\mathcal{V}ec(\mathbb{Z}_2)$}{}}

When $A_1=1$ with $f_{111}=1$, and $A_2=N_0\oplus N_1$ with $g_{N_i N_j N_K}=\delta_{N_i N_j N_k}$ are both Frobenius algebras. Then the multiplication after basis transformation is
\begin{equation}
\begin{aligned}
    &g^{N_0N_0N_0}_{111}=g^{N_0N_0N_0}_{1\psi\psi}=g^{N_0N_0N_0}_{\psi\psi1}=g^{N_0N_0N_0}_{\psi 1\psi}=1,\\
    &g^{N_0N_1N_1}_{111}=g^{N_0N_1N_1}_{1\psi\psi}=1,\ g^{N_0N_1N_1}_{\psi 1 \psi}=g^{N_0N_1N_1}_{\psi\psi1}=-1,
\end{aligned}
\end{equation}
with cyclic symmetry. 
When $\eta_{1N_0}=\eta_{1N_1}=\frac{1}{\sqrt{2}}$, the two $A_1\overset{\eta}{-}A_2$-bimodules are the same as Eq. \eqref{eq: AAbimoduleM0-A=1} and \eqref{eq: AAbimoduleM1-A=1}. Nevertheless, when $A_1=1 \oplus \psi$ with $f_{abc}=\delta_{abc}$, and $\eta_{1 N_0}=\eta_{1 N_1}=\eta_{\psi N_0}=\eta_{\psi N_1}=\frac{1}{2}$, there are two $\AAeta$-bimodules which are the same as Eq. \eqref{eq: ABbimodulesM0'} and Eq. \eqref{eq: ABbimodulesM1'}. 

When $A=1$ with $f_{111}=1$, and $A_2=N_0$ with $g_{N_0N_0N_0}=1$ are both Frobenius algebras. Then the multiplication after basis transformation is
\begin{equation}
    g^{N_0N_0N_0}_{111}=g^{N_0N_0N_0}_{1\psi\psi}=g^{N_0N_0N_0}_{\psi\psi1}=g^{N_0N_0N_0}_{\psi 1\psi}=1.
\end{equation} 
Then the two $A_1\overset{\eta}{-}A_2$-bimodules are the same as Eq. \eqref{eq: ABbimodulesM0} and \eqref{eq: ABbimodulesM1}.

When $A_1=1\oplus \psi$ with $f_{abc}=\delta_{abc}$, and $A_2=N_0$ with $g_{N_0N_0N_0}=1$ are both Frobenius algebras. The two $A_1\overset{\eta}{-}A_2$-bimodules are the same as Eq. \eqref{eq: AAbimodulesM0-A=1+psi} and \eqref{eq: AAbimodulesM1-A=1+psi}.

When $A_1=1\oplus \psi$ with $f_{abc}=\delta_{abc}$, $A_2=N_0\oplus N_1$ with $g_{N_i N_j N_K}=\delta_{N_i N_j N_k}$. Then $A_2$ has different multiplication after basis transformation as shown below:
\begin{equation}
    \begin{aligned}
        \Tilde{g}_{111}^{N_0 N_0 N_0} &= \Tilde{g}_{1\psi\psi}^{N_0 N_0 N_0} = \Tilde{g}_{\psi 1 \psi}^{N_0 N_0 N_0}= \Tilde{g}_{\psi\psi 1}^{N_0 N_0 N_0} = 1 \\
        \Tilde{g}_{111}^{N_0 N_1 N_1} &= \Tilde{g}_{1\psi\psi}^{N_0 N_1 N_1} = 1, \  \Tilde{g}_{\psi\psi 1}^{N_0 N_1 N_1}= i, \ \Tilde{g}_{\psi 1 \psi}^{N_0 N_1 N_1} = -i.
    \end{aligned}
\end{equation}

The gluing function is $\eta_{1N_0}=\eta_{\psi N_1}=\frac{1}{\sqrt{2}}$.
Then there are four irreducible $A_1\overset{\eta}{-}A_2$-bimodules: 
\begin{itemize}
    \item $M_0={1\oplus\psi}$ 
    \begin{equation}
        \begin{aligned}
            P^{11}_{111}&=1 \quad P^{11}_{\psi\psi\psi}=1 \quad P^{1\psi}_{1\psi\psi} = -i \quad P^{1\psi}_{\psi 1 1} = i \\
            P^{\psi 1}_{\psi \psi 1} &= 1 \quad 
            P^{\psi 1}_{1 1 \psi} = 1 \quad P^{\psi\psi}_{1\psi 1}=1 \quad P^{\psi\psi}_{\psi 1\psi}=-1 
        \end{aligned}
        \label{eq: ANbimodulesM0}
    \end{equation}

        \item $M_1={1\oplus\psi}$ 
    \begin{equation}
        \begin{aligned}
            P^{11}_{111}&=1 \quad P^{11}_{\psi\psi\psi}=1 \quad P^{1\psi}_{1\psi\psi} = i \quad P^{1\psi}_{\psi 1 1} = -i \\
            P^{\psi 1}_{\psi \psi 1 } &= -1 \quad 
            P^{\psi 1}_{1 1 \psi} = -1 \quad P^{\psi\psi}_{1\psi 1}=1 \quad P^{\psi\psi}_{\psi 1\psi}=-1 
        \end{aligned}
        \label{eq: ANbimodulesM1}
    \end{equation}
    
    \item $M_2={1\oplus\psi}$ 
    \begin{equation}
        \begin{aligned}
            P^{11}_{111}&=1 \quad P^{11}_{\psi\psi\psi}=1 \quad P^{1\psi}_{1\psi\psi} = -i \quad P^{1\psi}_{\psi 1 1} = i \\
            P^{\psi 1}_{\psi \psi 1}& = -1 \quad 
            P^{\psi 1}_{1 1 \psi} = -1 \quad
            P^{\psi\psi}_{1\psi 1}=-1 \quad P^{\psi\psi}_{\psi 1\psi}=1
        \end{aligned}
        \label{eq: ANbimodulesM2}
    \end{equation}

        \item $M_3={1\oplus\psi}$ 
    \begin{equation}
        \begin{aligned}
            P^{11}_{111}&=1 \quad P^{11}_{\psi\psi\psi}=1 \quad P^{1\psi}_{1\psi\psi} = i \quad P^{1\psi}_{\psi 1 1} = -i \\
            P^{\psi 1}_{\psi \psi 1} &= 1 \quad 
            P^{\psi 1}_{1 1 \psi} = 1 \quad
            P^{\psi\psi}_{1\psi 1}=-1 \quad P^{\psi\psi}_{\psi 1\psi}=1
        \end{aligned}
        \label{eq: ANbimodulesM3}
    \end{equation}
\end{itemize}

\subsection{\texorpdfstring{$\AAeta$}{}-bimodules for the Ising LW model \label{Appendix: A1-bimodules-Ising}}

When $A=1$ and $\eta_{11}=1$, there are three irreducible $A$-bimodules: 
\begin{itemize}
    \item $M_0={1}$ 
        \begin{equation}
         P^{11}_{111}=1 \end{equation}
    \item $M_1={\psi}$ 
        \begin{equation}
        P^{11}_{\psi \psi \psi}=1
        \end{equation}
    \item $M_2={\sigma}$ 
        \begin{equation}
          P^{11}_{\sigma \sigma \sigma}=1
        \end{equation}
\end{itemize}

When $A=1\oplus \psi$, and $\eta_{ab}=\delta_{a\in \{1, \psi\}} \delta_{b\in \{1, \psi\}}$ is separable, there are also three irreducible $\AAeta$-bimodules. 

\begin{itemize}
    \item $M_0={1\oplus\psi}$ 
    \begin{equation}
        \begin{aligned}
            P^{11}_{111}&=1 \quad P^{11}_{\psi\psi\psi}=1 \quad P^{1\psi}_{1\psi\psi}=1 \quad P^{1\psi}_{\psi 1 1}=1 \\
            P^{\psi 1}_{\psi \psi 1}&=1 \quad 
            P^{\psi 1}_{1 1 \psi}=1 \quad P^{\psi\psi}_{1\psi 1}=1 \quad P^{\psi\psi}_{\psi 1\psi}=1 
        \end{aligned}
    \end{equation}
    
    \item $M_1={1\oplus\psi}$ 
    \begin{equation}
        \begin{aligned}
            P^{11}_{111}&=1 \quad P^{11}_{\psi\psi\psi}=1 \quad P^{1\psi}_{1\psi\psi}=-1 \quad P^{1\psi}_{\psi 1 1}=-1 \\ 
            P^{\psi 1}_{\psi \psi 1}&=1 \quad 
            P^{\psi 1}_{1 1 \psi}=1 \quad
            P^{\psi\psi}_{1\psi 1}=-1 \quad P^{\psi\psi}_{\psi 1\psi}=-1
        \end{aligned}
    \end{equation}
            
    \item $M_2={\sigma_1\oplus\sigma_2}$
    \begin{equation}
        \begin{aligned}
            P^{11}_{\sigma_1 \sigma_1 \sigma_1}&=\frac{1}{4} \quad 
            P^{11}_{\sigma_1 \sigma_2 \sigma_1}=\frac{1}{4} \quad 
            P^{11}_{\sigma_2\sigma_1 \sigma_2}=\frac{1}{4} \quad 
            P^{11}_{\sigma_2 \sigma_2 \sigma_2}=\frac{1}{4} \\          
            P^{1\psi}_{\sigma_1 \sigma_1\sigma_1}&=\frac{1}{4} \quad P^{1\psi}_{\sigma_1\sigma_2\sigma_1}=\frac{1}{4} \quad P^{1\psi}_{\sigma_2\sigma_1\sigma_2}=-\frac{1}{4} \quad P^{1\psi}_{\sigma_2\sigma_2\sigma_2}=-\frac{1}{4} \\
            P^{\psi 1}_{\sigma_1\sigma_1\sigma_2}&=\frac{1}{4} \quad 
            P^{\psi 1}_{\sigma_2 \sigma_1 \sigma_1}=\frac{1}{4} \quad 
            P^{\psi 1}_{\sigma_2 \sigma_2\sigma_1}=\frac{1}{4} \quad 
            P^{\psi 1}_{\sigma_1 \sigma_2\sigma_2}=\frac{1}{4} \\
            P^{\psi\psi}_{\sigma_1\sigma_1\sigma_2}&=-\frac{1}{4} \quad P^{\psi\psi}_{\sigma_1\sigma_2\sigma_2}=-\frac{1}{4} \quad P^{\psi\psi}_{\sigma_2\sigma_1\sigma_1}=\frac{1}{4} \quad
            P^{\psi\psi}_{\sigma_2\sigma_2\sigma_1}=\frac{1}{4} \\
        \end{aligned}
    \end{equation}

\end{itemize}

When $A=1\oplus \psi$, and $\eta_{ab}=\delta_{ab}, \forall a,b\in A$, there are six irreducible $\AAeta$-bimodules for $A=1\oplus \psi$:
\begin{itemize}
    \item $M_0=1$ 
    \begin{equation}
            P^{11}_{111}=1 \quad P^{\psi\psi}_{1\psi 1}=1 
    \end{equation}
            
    \item $M_1=\psi$
  \begin{equation}
            P^{11}_{\psi \psi\psi}=1 \quad P^{\psi\psi}_{\psi 1 \psi}=1 
    \end{equation}

    \item $M_2=1$
  \begin{equation}
            P^{11}_{111}=1 \quad P^{\psi\psi}_{1\psi 1}=-1 
    \end{equation}
    
    \item $M_3=\psi$
  \begin{equation}
            P^{11}_{\psi \psi\psi}=1 \quad P^{\psi\psi}_{\psi 1 \psi}=-1 
    \end{equation}

    \item $M_4=\sigma$
  \begin{equation}
            P^{11}_{\sigma \sigma \sigma}=1 \quad P^{\psi\psi}_{\sigma \sigma \sigma}= i 
    \end{equation}
    
    \item $M_5=\sigma$
  \begin{equation}
            P^{11}_{\sigma \sigma \sigma}=1 \quad P^{\psi\psi}_{\sigma \sigma \sigma}= -i
    \end{equation}
    
\end{itemize}



\bibliographystyle{JHEP}
\bibliography{biblio.bib}

\providecommand{\href}[2]{#2}\begingroup\raggedright\begin{thebibliography}{10}

\bibitem{hu2018boundary}
Y.~Hu, Z.-X.~Luo, R.~Pankovich, Y.~Wan and Y.-S.~Wu, \emph{Boundary hamiltonian theory for gapped topological phases on an open surface}, {\emph{Journal of High Energy Physics} {\bfseries 2018} (2018) 1}.

\bibitem{kapustin2011topological}
A.~Kapustin and N.~Saulina, \emph{Topological boundary conditions in abelian chern--simons theory}, {\emph{Nuclear Physics B} {\bfseries 845} (2011) 393}.

\bibitem{fuchs2013bicategories}
J.~Fuchs, C.~Schweigert and A.~Valentino, \emph{Bicategories for boundary conditions and for surface defects in 3-d tft}, {\emph{Communications in Mathematical Physics} {\bfseries 321} (2013) 543}.

\bibitem{lan2015gapped}
T.~Lan, J.C.~Wang and X.-G.~Wen, \emph{Gapped domain walls, gapped boundaries, and topological degeneracy}, {\emph{Physical review letters} {\bfseries 114} (2015) 076402}.

\bibitem{Kitaev-Kong-2012}
A.~Kitaev and L.~Kong, \emph{{Models for Gapped Boundaries and Domain Walls}}, \href{https://doi.org/https://doi.org/10.1007/s00220-012-1500-5}{\emph{Communications in Mathematical Physics} {\bfseries 313} (2012) }.

\bibitem{jia2023boundary}
Z.~Jia, D.~Kaszlikowski and S.~Tan, \emph{Boundary and domain wall theories of 2d generalized quantum double model}, {\emph{Journal of High Energy Physics} {\bfseries 2023} (2023) 1}.

\bibitem{zhao2023characteristic}
Y.~Zhao, S.~Huang, H.~Wang, Y.~Hu and Y.~Wan, \emph{Characteristic properties of a composite system of topological phases separated by gapped domain walls via an exactly solvable hamiltonian model}, {\emph{SciPost Physics Core} {\bfseries 6} (2023) 076}.

\bibitem{christian2023lattice}
J.~Christian, D.~Green, P.~Huston and D.~Penneys, \emph{A lattice model for condensation in levin-wen systems}, {\emph{Journal of High Energy Physics} {\bfseries 2023} (2023) 1}.

\bibitem{beigi2011quantum}
S.~Beigi, P.W.~Shor and D.~Whalen, \emph{The quantum double model with boundary: condensations and symmetries}, {\emph{Communications in mathematical physics} {\bfseries 306} (2011) 663}.

\bibitem{Cong-TQC-2017}
I.~Cong and Z.~Wang, \emph{{Topological quantum computation with gapped boundaries and boundary defects}}, \href{https://doi.org/10.1090/conm/747/15043}{\emph{Topological Phases of Matter and Quantum Computation} (2017) }.

\bibitem{cong2017universal}
I.~Cong, M.~Cheng and Z.~Wang, \emph{Universal quantum computation with gapped boundaries}, {\emph{Physical Review Letters} {\bfseries 119} (2017) 170504}.

\bibitem{Bais2009}
F.A.~Bais and J.K.~Slingerland, \emph{{Condensate-induced transitions between topologically ordered phases}}, \href{https://doi.org/10.1103/PhysRevB.79.045316}{\emph{Physical Review B} {\bfseries 79} (2009) }.

\bibitem{Bais2009-prl}
F.A.~Bais, J.K.~Slingerland and S.M.~Haaker, \emph{{Theory of topological edges and domain walls}}, \href{https://doi.org/10.1103/PhysRevLett.102.220403}{\emph{Physical Review Letters} {\bfseries 102} (2009) }.

\bibitem{Kong-2014}
L.~Kong, \emph{{Anyon condensation and tensor categories}}, \href{https://doi.org/10.1016/j.nuclphysb.2014.07.003}{\emph{Nuclear Physics B} {\bfseries 886} (2014) }.

\bibitem{neupert2016boson}
T.~Neupert, H.~He, C.~Von~Keyserlingk, G.~Sierra and B.A.~Bernevig, \emph{Boson condensation in topologically ordered quantum liquids}, {\emph{Physical Review B} {\bfseries 93} (2016) 115103}.

\bibitem{hu2022anyon}
Y.~Hu, Z.~Huang, L.-Y.~Hung and Y.~Wan, \emph{Anyon condensation: coherent states, symmetry enriched topological phases, goldstone theorem, and dynamical rearrangement of symmetry}, {\emph{Journal of High Energy Physics} {\bfseries 2022} (2022) 1}.

\bibitem{kesselring2024anyon}
M.S.~Kesselring, J.C.~Magdalena de~la Fuente, F.~Thomsen, J.~Eisert, S.D.~Bartlett and B.J.~Brown, \emph{Anyon condensation and the color code}, {\emph{PRX Quantum} {\bfseries 5} (2024) 010342}.

\bibitem{zhao2024symmetry}
Y.~Zhao, H.~Wang, Y.~Hu and Y.~Wan, \emph{Symmetry fractionalized (irrationalized) fusion rules and two domain-wall verlinde formulae}, {\emph{Journal of High Energy Physics} {\bfseries 2024} (2024) 1}.

\bibitem{zhao2024nonabelian}
Y.~Zhao and Y.~Wan, \emph{Nonabelian anyon condensation in 2+ 1d topological orders: A string-net model realization}, {\emph{arXiv preprint arXiv:2409.05852} (2024) }.

\bibitem{bombin2010topological}
H.~Bomb{\'\i}n, \emph{Topological order with a twist: Ising anyons from an abelian model}, {\emph{Physical review letters} {\bfseries 105} (2010) 030403}.

\bibitem{buerschaper2013electric}
O.~Buerschaper, M.~Christandl, L.~Kong and M.~Aguado, \emph{Electric--magnetic duality of lattice systems with topological order}, {\emph{Nuclear Physics B} {\bfseries 876} (2013) 619}.

\bibitem{wang2020electric}
H.~Wang, Y.~Li, Y.~Hu and Y.~Wan, \emph{Electric-magnetic duality in the quantum double models of topological orders with gapped boundaries}, {\emph{Journal of High Energy Physics} {\bfseries 2020} (2020) 1}.

\bibitem{hu2020electric}
Y.~Hu and Y.~Wan, \emph{Electric-magnetic duality in twisted quantum double model of topological orders}, {\emph{Journal of High Energy Physics} {\bfseries 2020} (2020) 1}.

\bibitem{huston2023composing}
P.~Huston, F.~Burnell, C.~Jones and D.~Penneys, \emph{Composing topological domain walls and anyon mobility}, {\emph{SciPost Physics} {\bfseries 15} (2023) 076}.

\bibitem{hu2018full}
Y.~Hu, N.~Geer and Y.-S.~Wu, \emph{Full dyon excitation spectrum in extended levin-wen models}, {\emph{Physical Review B} {\bfseries 97} (2018) 195154}.

\bibitem{Hung-Wan-ADE-2015}
L.Y.~Hung and Y.~Wan, \emph{{Generalized ADE classification of topological boundaries and anyon condensation}}, \href{https://doi.org/10.1007/JHEP07(2015)120}{\emph{Journal of High Energy Physics} (2015) }.

\bibitem{kock2004frobenius}
J.~Kock, \emph{Frobenius Algebras and 2-D Topological Quantum Field Theories}, London Mathematical Society Student Texts, Cambridge University Press (2003).

\bibitem{wang2022extend}
H.~Wang, Y.~Hu and Y.~Wan, \emph{Extend the levin-wen model to two-dimensional topological orders with gapped boundary junctions}, {\emph{Journal of High Energy Physics} {\bfseries 2022} (2022) 1}.

\bibitem{cheng2023precision}
G.~Cheng, L.~Chen, Z.-C.~Gu and L.-Y.~Hung, \emph{Precision reconstruction of rational cft from exact fixed point tensor network}, {\emph{arXiv e-prints} (2023) arXiv}.

\bibitem{chen2024cft}
L.~Chen, K.~Ji, H.~Zhang, C.~Shen, R.~Wang, X.~Zeng et~al., \emph{Cft d from tqft d+ 1 via holographic tensor network, and precision discretization of cft 2}, {\emph{Physical Review X} {\bfseries 14} (2024) 041033}.

\bibitem{levin2005string}
M.A.~Levin and X.-G.~Wen, \emph{String-net condensation: A physical mechanism for topological phases}, {\emph{Physical Review B} {\bfseries 71} (2005) 045110}.

\bibitem{zhao2024noninvertible}
Y.~Zhao and Y.~Wan, \emph{Noninvertible gauge symmetry in (2+ 1) d topological orders: A string-net model realization}, {\emph{arXiv preprint arXiv:2408.02664} (2024) }.

\end{thebibliography}\endgroup



\end{document}